\documentclass[5p]{elsarticle}

\usepackage[utf8]{inputenc}
\usepackage[T1]{fontenc}
\usepackage{lmodern}
\usepackage{amsmath,amssymb}
\usepackage{graphicx}
\usepackage{booktabs}
\usepackage{multirow}
\usepackage{array}
\usepackage[font=small,labelfont=bf]{caption}
\usepackage{microtype}
\usepackage{placeins}
\usepackage{xcolor}
\usepackage[colorlinks=true,linkcolor=blue!50!black,citecolor=blue!50!black,urlcolor=blue!50!black]{hyperref}
\usepackage[capitalise,nameinlink]{cleveref}
\usepackage{orcidlink}

\crefformat{appendix}{#2#1#3}\Crefformat{appendix}{#2#1#3}
\crefrangeformat{appendix}{#3#1#4 to #5#2#6}\Crefrangeformat{appendix}{#3#1#4 to #5#2#6}
\crefmultiformat{appendix}{#2#1#3}{ and #2#1#3}{, #2#1#3}{ and #2#1#3}\Crefmultiformat{appendix}{#2#1#3}{ and #2#1#3}{, #2#1#3}{ and #2#1#3}
\crefformat{subappendix}{#2#1#3}\Crefformat{subappendix}{#2#1#3}
\crefrangeformat{subappendix}{#3#1#4 to #5#2#6}\Crefrangeformat{subappendix}{#3#1#4 to #5#2#6}
\crefmultiformat{subappendix}{#2#1#3}{ and #2#1#3}{, #2#1#3}{ and #2#1#3}\Crefmultiformat{subappendix}{#2#1#3}{ and #2#1#3}{, #2#1#3}{ and #2#1#3}
\crefformat{subsubappendix}{#2#1#3}\Crefformat{subsubappendix}{#2#1#3}
\crefrangeformat{subsubappendix}{#3#1#4 to #5#2#6}\Crefrangeformat{subsubappendix}{#3#1#4 to #5#2#6}
\crefmultiformat{subsubappendix}{#2#1#3}{ and #2#1#3}{, #2#1#3}{ and #2#1#3}\Crefmultiformat{subsubappendix}{#2#1#3}{ and #2#1#3}{, #2#1#3}{ and #2#1#3}

\graphicspath{{figures/}}
\newcommand{\ttbar}{t\bar{t}}
\newcommand{\mt}{m_t}
\newcommand{\mtmc}{m_t^{\mathrm{MC}}}
\newcommand{\dmtmc}{\Delta m_t^{\mathrm{MC}}}
\newcommand{\mW}{m_W}
\newcommand{\mlb}{m_{\ell b}}
\newcommand{\mlbmm}{m_{\ell b}^{\mathrm{minimax}}}
\newcommand{\pT}{p_{\mathrm{T}}}
\newcommand{\Ninb}{N_{\text{in-}b}}
\newcommand{\ptinb}{\bar{p}_{\mathrm{T},\text{in-}b}}
\newcommand{\NUE}{N_{\mathrm{UE}}}
\newcommand{\SUE}{S_{\mathrm{UE}}}
\newcommand{\chimin}{\chi^2_{\min}}
\newcommand{\GeV}{\,\mathrm{GeV}}
\newcommand{\MeV}{\,\mathrm{MeV}}
\newcommand{\TeV}{\,\mathrm{TeV}}
\newcommand{\pb}{\,\mathrm{pb}}
\newcommand{\ifb}{\,\mathrm{fb}^{-1}}
\newcommand{\sqrts}{\sqrt{s}}
\newcommand{\emu}{e\mu}
\newcommand{\mtag}[1]{\texttt{#1}}

\newcommand{\orcidAB}{0009-0002-7563-1176}
\newcommand{\orcidFF}{0000-0002-6423-7213}

\journal{Nuclear Physics B}
\makeatletter
\def\ps@pprintTitle{\let\@oddhead\@empty\let\@evenhead\@empty
  \def\@oddfoot{\reset@font\hfil\thepage\hfil}\let\@evenfoot\@oddfoot}
\makeatother

\begin{document}

\begin{frontmatter}

\title{Disentangling Colour Reconnection and Top-Quark Mass Using Charged-Particle Flow}

\author[mvu]{Ahmed Bellagroudi\,\orcidlink{\orcidAB}\corref{cor1}}
\makeatletter\global\@eadauthor={Ahmed Bellagroudi}\makeatother 
\ead{ahmed.bellagroudi@cern.ch}
\author[mvu]{Farida Fassi\,\orcidlink{\orcidFF}}
\makeatletter\global\@eadauthor={Farida Fassi}\makeatother
\ead{farida.fassi@cern.ch}
\cortext[cor1]{Corresponding author}
\affiliation[mvu]{organization={Laboratory of Condensed Matter and Interdisciplinary Sciences, Unit\'e de Recherche Labellis\'ee CNRST (URL-CNRST), Faculty of Sciences, Mohammed V University in Rabat},
  city={Rabat},
  postcode={1014},
  country={Morocco}}

\begin{abstract}
Colour reconnection (CR) is one of the largest modelling uncertainties in direct measurements of the top-quark mass. Using \textsc{Pythia}~8.318 at particle level, we ask whether charged particles in and around the $b$-jets of dilepton $\ttbar$ events can measure CR in the same events that measure the mass. Different CR models shift the mass extracted from the lepton--$b$-jet invariant mass $\mlb$ by $-0.73$ to $+0.28\GeV$, and the shift caused by CR involving the top-decay products follows the change in the charged particles inside the $b$-jets. We therefore fit the mass and these charged-particle observables together, with the CR strength as a free parameter. In 24 blind test samples this nearly halves the mass error, from $0.18$ to $0.096\GeV$, with coverage close to nominal (16/24 at 68\%, 24/24 at 95\%). Outside the trained models, a goodness-of-fit test flags CR switched off and complete CMS tunes, but not every alternative model, and it cannot correct the CMS tunes: they shift the mass by $+0.43$ to $+0.67\GeV$, so the calibration holds only within one tune. The mass and charged-particle responses to CR are confirmed at next-to-leading order. Changes to the recoil scheme in top decay and to $b$-quark fragmentation shift the mass by similar amounts ($+0.25$ and $-0.16\GeV$) and are not corrected. The ATLAS measurement of charged particles outside jets is sensitive to this CR, but the CMS CP5 tune does not describe it ($\chi^2=93$ for 6 degrees of freedom), so the data only prefer a small CR strength and do not constrain it.

\end{abstract}


\end{frontmatter}

\section{Introduction}
\label{sec:intro}

The top mass is now measured with a precision of about 0.2\%, and the remaining uncertainty is dominated by modelling rather than statistics. The ATLAS and CMS Run~1 combination reaches a total uncertainty of about $0.33\GeV$~\cite{ATLAS:2024dxp}. Direct measurements extract the mass parameter of a Monte Carlo (MC) generator, $\mtmc$. Its relation to a field-theory mass scheme is a separate, active question~\cite{Hoang:2008xm,Nason:2017cxd,Hoang:2020iah}. In this paper we will address the first step only: how well $\mtmc$ can be extracted when the hadronisation model itself is uncertain.

Colour reconnection (CR) rearranges the colour flow between partons from the top decays, the beam remnants and multiple parton interactions (MPI). Because the $b$-quarks from top decay are colour-connected to the rest of the event, CR changes the energy and particle content of $b$-jets. Different CR models move the reconstructed mass by several hundred MeVs~\cite{Argyropoulos:2014zoa,Christiansen:2015yqa}. Argyropoulos and Sj\"ostrand, who introduced the top-specific CR models used here, also proposed charged-particle observables in $\ttbar$ events, such as the multiplicity and scalar $\pT$ sum outside jets, as future constraints on these models~\cite{Argyropoulos:2014zoa}. Hadronic mass observables also require the $b$-fragmentation and hadronisation parameters to be known precisely~\cite{Corcella:2017rpt}. Experiments currently assign this as a systematic uncertainty from a spread of models, such as the CMS CP5-CR tunes~\cite{CMS:2022awf}.

The dilepton channel is cleaner than lepton+jets but has no hadronic $W$ boson. There is therefore no in-situ handle on the $b$-jet response, and CR enters the mass almost entirely through the $b$-jets and their surroundings. ATLAS has measured charged particle distributions in $\ttbar$ events that are sensitive to CR~\cite{ATLAS:2022zbu}, ATLAS has measured charged particle distributions in $\ttbar$ events that are sensitive to CR~\cite{ATLAS:2022zbu}. The measurement also compared models with different strengths of top-specific CR and found the scalar $\pT$ sum, measured in bins of the charged multiplicity, to be the most sensitive observable. These data have so far been used to test models, not to constrain the top mass. Other mass observables reduce the dependence on jets, using only leptonic quantities~\cite{Frixione:2014ala} or the invariant mass of a lepton and a charged particle secondary vertex~\cite{CMS:2016iru}. Energy correlator observables have been proposed as theoretically cleaner mass probes~\cite{Holguin:2022epo,Holguin:2023bjf,Holguin:2024tkz}, their CR sensitivity has been studied mainly through generic CR variants.

Fitting the mass together with modelling nuisance parameters is not new in itself, inference of the top mass from event ensembles with simultaneous nuisance constraints has been studied with machine learning methods~\cite{Flesher:2020kuy}, CMS has constrained modelling nuisance parameters in situ with several observables in a profile-likelihood mass measurement~\cite{CMS:2023ebf}, and CMS derived CR tunes partly to assess their top mass impact~\cite{CMS:2022awf}. Charged particle observables are therefore not new as a probe of top-specific CR. Calibrating generator parameters in situ, with observables that depend only weakly on the top mass, was proposed and quantified by Corcella, Franceschini and Kim for shower and hadronisation parameters~\cite{Corcella:2017rpt}, colour reconnection was not among the parameters they varied. We specialise that strategy to a continuous top-specific CR strength. What we add is the connection of charged particle observables to a calibrated top mass inference: a charged particle flow response model built to separate the top mass from top-specific CR, prospective blinded tests of its bias, coverage and failure detection, and a demonstration of where and why the flow to mass relation breaks down, in particular when the generator tune changes.
The approach follows the logic of the in-situ jet energy scale in lepton+jets measurements, where the hadronic $W$ mass constrains the light jet response in the same events that measure $\mt$. In the dilepton channel no such resonance is available for the $b$-jets. We ask whether the charged particle content of the $b$-jets can play that role for the CR induced part of their response.

Concretely: can charged particle features of the same dilepton events, determine the CR strength and correct the mass bias, and does the procedure stay honest when the true model lies outside the family it was trained on? The paper proceeds as follows.
\begin{enumerate}
  \item \Cref{sec:theory} explains how the $\mlb$ fit depends on the $b$-jet energy, and shows that the mass shift caused by top-specific CR can be predicted from the change in the charged particle $\pT$ inside the $b$-jet cones.
  \item \Cref{sec:setup,sec:mlb,sec:fingerprint} measure the $\mlb$ fit bias for CR, MPI and $b$-fragmentation variations, and build a nine-feature charged particle fingerprint that separates CR from $b$-fragmentation.
  \item \Cref{sec:inference} sets up a joint mass+flow likelihood with a continuous CR strength and profiled nuisance parameters, and \cref{sec:blind} tests it in three blind tests: inside the trained family, outside of it, and with independently tuned CMS models.
  \item \Cref{sec:nlo} validates the calibration and the responses in a matched next-to-leading-order (NLO) chain.
  \item \Cref{sec:recoil} studies the top-decay recoil scheme and $b$-quark fragmentation, which a CR flow correction does not capture.
  \item \Cref{sec:rivet} compares with the ATLAS measurement of charged particles outside jets, after validating its published covariance matrices.
\end{enumerate}
\Cref{tab:summary} collects the main results.

\begin{table*}[tbp]
  \centering
  \caption{Main results. RMS: root mean square, FSR: final-state radiation. LO: leading order (\textsc{Pythia}~8.318, Monash unless stated), NLO: next-to-leading order, MG5\_aMC@NLO with MC@NLO-matched \textsc{Pythia} CP5. Mass shifts refer to the $\mlb$ template fit.}
  \label{tab:summary}
  \footnotesize
  \renewcommand{\arraystretch}{1.3}
  \begin{tabular}{p{0.30\textwidth}p{0.62\textwidth}}
    \toprule
    Study & Result \\
    \midrule
    CR models, LO (\cref{sec:mlb}) & Fit moves by $-0.73$ to $+0.28\GeV$ across standard CR models \\
    Blind test in family, LO (\cref{sec:val24}) & RMS mass error $0.096\GeV$, against $0.18\GeV$ for a mass-only fit covering the CR family, coverage 16/24 (68\%) and 24/24 (95\%) \\
    Blind tests out of family, LO (\cref{sec:oof12,sec:cms7}) & Goodness-of-fit test flags CR off and the CMS tunes, not every alternative model, CMS tunes shift the fit by $+0.43$ to $+0.67\GeV$ and the fallback interval misses 5 of 6: the calibration is tune specific \\
    NLO validation (\cref{sec:nlo}) & Closure residual $-0.18\pm0.12\GeV$, responses to top-specific CR ($-0.48\pm0.11\GeV$) and FSR ($-0.61\pm0.11\GeV$) agree with LO \\
    Recoil and fragmentation, NLO (\cref{sec:recoil}) & Recoil-to-$b$ $+0.25\pm0.10\GeV$, Bowler $r_b=1.05$ $-0.16\pm0.11\GeV$, not captured by a CR flow correction \\
    ATLAS data, 13\,TeV (\cref{sec:rivet}) & Sensitive to top-specific CR (expected $\Delta\chi^2=24$), CP5 fails ($\chi^2_{\min}=93/6$), preference $\hat s=0.036$ within CP5, mass translation about $-0.1\GeV$ (diagnostic) \\
    \bottomrule
  \end{tabular}
\end{table*}

The study is at particle level with one generator, at leading order (LO) for the blinded tests and at next-to-leading order for the response validation. It is a proof of principle for the method and its validation protocol, not an estimate of the uncertainty in any experimental measurement. Its central lesson is that charged particle flow can constrain the CR induced mass bias within a specified model family, but good in family precision and coverage do not guarantee a reliable calibration across independently tuned generators.

\section{Theoretical framework}
\label{sec:theory}

This section sets out the quantum chromodynamics (QCD) basis of the problem: why colour reconnection biases a kinematic top-mass estimator and why the bias is visible in the charged particles of the $b$-jets. Where possible each statement is checked against the samples of \cref{sec:setup}. The colour flow and timing of top decay, heavy-quark fragmentation, and the condition under which mass and CR can be separated in the same events (identifiability) are discussed in \cref{app:theory}.

\subsection{The generator mass parameter}
\label{sec:theory-mass}

The pole mass of the top quark cannot be defined more precisely than about $\Lambda_{\mathrm{QCD}}$, a few hundred MeV, because of long-distance QCD effects (the infrared renormalon). Other mass definitions, such as the $\overline{\mathrm{MS}}$ and MSR masses, do not have this limitation~\cite{Hoang:2020iah}. Direct reconstruction does not measure any of these, it measures the value of $\mtmc$ at which the simulation best reproduces the data. Its relation to a field theory mass depends on the parton shower cutoff and on the non-perturbative model, and is the subject of dedicated calibration work~\cite{Hoang:2008xm,Nason:2017cxd,Hoang:2020iah}.

We do not address that relation. We treat $\mtmc$ as a parameter of a fixed generator setup and ask a prior question: by how much does the extracted $\mtmc$ move when only the colour-reconnection model changes, and can that motion be measured in situ? If changing only the CR model moves the extracted mass, then the same event kinematics correspond to a different value of the generator's top-mass parameter. A direct measurement calibrated with one CR model would then be biased if the true CR differs.

\subsection{QCD factorisation for an unstable coloured particle}
\label{sec:theory-qcd}

Event generators split a $\ttbar$ event into separate steps: the parton densities of the colliding protons, the hard production of the top pair, the top decays, a parton shower that adds radiation down to about $1\GeV$, and a hadronisation model below that scale. For the decays they use the narrow-width approximation, which treats each top quark as exactly on its mass shell by replacing the Breit--Wigner propagator with a delta function,
\begin{equation}
  \frac{1}{(p^2-\mt^2)^2+\mt^2\Gamma_t^2} \,\longrightarrow\, \frac{\pi}{\mt\Gamma_t}\,\delta(p^2-\mt^2),
  \label{eq:nwa}
\end{equation}
with corrections of relative order $\Gamma_t/\mt\approx0.8\%$. Beyond this approximation, two features of QCD limit how cleanly production and decay can be separated.

\paragraph{Soft gluons and the top width} Gluons radiated in the production and in the decay interfere only if their energy is too low to resolve the time between the two stages. For a gluon of energy $\omega$ the interference is weighted by~\cite{Orr:1992gl}
\begin{equation}
  \chi(\omega) = \frac{\Gamma_t^2}{\Gamma_t^2+\omega^2},
  \label{eq:chi}
\end{equation}
which for $\Gamma_t=1.40\GeV$ gives $\chi=0.66$ at $\omega=1\GeV$ and $0.07$ at $\omega=5\GeV$. Hard gluons are therefore radiated independently by the production and the decay, which justifies showering the two separately. Only soft gluons, with $\omega\lesssim\Gamma_t$, see the colour structure of the whole event, and this energy range overlaps the non-perturbative region.

\paragraph{Colour suppression} Generators assign colour connections in the limit of a large number of colours, $N_c\to\infty$, where each parton has a definite colour partner. Corrections to this assignment are formally of order $1/N_c^2\approx11\%$, although this is only a counting estimate, not a general prediction of the size of CR effects. In perturbation theory, colour connections between two separate colour-singlet systems, such as the two $W$ bosons in $W^+W^-$ production, can be rearranged only at second order in $\alpha_s$, and the effect is further reduced by colour factors. Larger effects need non-perturbative reconnection at later times, during hadronisation~\cite{Sjostrand:1993rb}. Colour reconnection therefore cannot be calculated from first principles. It has to be modelled and its strength measured, which motivates an in-situ constraint.

\subsection{Kinematics of the \texorpdfstring{$m_{\ell b}$}{mlb} observable}
\label{sec:theory-mlb}

For $t\to bW^+\to b\ell^+\nu$ with massless $b$ and $\ell$, evaluate $m_{\ell b}^2=2E_\ell E_b(1-\cos\theta_{\ell b})$ in the $W$ rest frame, where $E_\ell=\mW/2$ and $E_b=(\mt^2-\mW^2)/(2\mW)$. Writing $\theta^*$ for the angle between the charged lepton and the $W$ flight direction in the top rest frame, the $b$-quark moves along $-\hat{z}$ in the $W$ frame and
\begin{equation}
  m_{\ell b}^2 = \frac{\mt^2-\mW^2}{2}\,\big(1+\cos\theta^*\big),
  \qquad 0 \le m_{\ell b} \le \sqrt{\mt^2-\mW^2}.
  \label{eq:mlb-kin}
\end{equation}
The angular distribution is fixed by the $W$ helicity fractions,
\begin{equation}
  \begin{split}
  \frac{1}{\Gamma}\frac{d\Gamma}{d\cos\theta^*} &= \frac34 F_0\,(1-\cos^2\theta^*)\\
  &\quad + \frac38 F_L\,(1-\cos\theta^*)^2 + \frac38 F_R\,(1+\cos\theta^*)^2,
  \end{split}
  \label{eq:helicity}
\end{equation}
with $F_0=\mt^2/(\mt^2+2\mW^2)$, $F_L=2\mW^2/(\mt^2+2\mW^2)$ and $F_R=0$ at leading order. At $\mt=172.5\GeV$ and $\mW=80.385\GeV$ this gives $F_0=0.697$, $F_L=0.303$, an endpoint of $152.6\GeV$ and $\langle m_{\ell b}\rangle=94.1\GeV$.

Differentiating the mean of \cref{eq:mlb-kin} over \cref{eq:helicity}, including the mass dependence of $F_0$ and $F_L$, gives a parton-level slope $d\langle m_{\ell b}\rangle/d\mt=0.78$, or $0.75$ inside the $30$--$160\GeV$ window. The particle-level slope measured from the five nominal samples is $0.706\pm0.046$ (\cref{fig:calib}), consistent with the parton-level value. The residual dilution comes from wrong pairings, radiation outside the $R=0.4$ cones and neutrinos from semileptonic $b$ decays.

\subsection{How colour reconnection enters: a \texorpdfstring{$b$}{b}-jet response proxy}
\label{sec:theory-response}

At fixed decay angles, $m_{\ell b}^2\propto E_b$, so a fractional change of the reconstructed $b$-jet energy produces
\begin{equation}
  \frac{\delta m_{\ell b}}{m_{\ell b}} \simeq \frac12\,\frac{\delta E_b}{E_b}.
  \label{eq:jes}
\end{equation}
CR does not change the decay kinematics, it changes how much of the $b$-quark's energy ends up inside the cone. If CR acts mainly through the jet energy, \cref{eq:jes} should link the shift of $\langle\mlb\rangle$ to the shift of the energy in the cones. \Cref{tab:response} tests this using the charged particle $\pT$ summed over both cones, $S_{\text{in-}b}$ ($\langle S_{\text{in-}b}\rangle=101.2\GeV$ for the nominal sample), as a proxy for the jet energy. $S_{\text{in-}b}$ is not the full jet momentum, so applying \cref{eq:jes} to it gives only an approximate measure of the jet response, not a jet-energy calibration. The comparison below tests whether it accounts for the mass shift.

\begin{table*}[tbp]
  \centering
  \caption{Fractional shifts relative to the nominal sample (mean of two seeds, $\mtmc=172.5\GeV$, events in the $\mlb$ fit window): summed charged $\pT$ in the two $b$-jet cones, the prediction of \cref{eq:jes} from it, the observed shift of $\langle\mlb\rangle$, and the shift of the mean underlying-event (UE) multiplicity, i.e.\ of charged particles outside all jets ($\langle\NUE\rangle=45.2$ for the nominal). Model tags are defined in \cref{tab:models} (top: top-specific CR, bfrag: Bowler $b$-fragmentation, cr\_off: no CR, m0: MPI-based CR). Uncertainties are statistical.}
  \label{tab:response}
  \small
  \begin{tabular}{lcccc}
    \toprule
    Model & $\delta S_{\text{in-}b}/S_{\text{in-}b}$ [\%] & $\tfrac12\,\delta S_{\text{in-}b}/S_{\text{in-}b}$ [\%] & $\delta\langle\mlb\rangle/\langle\mlb\rangle$ [\%] & $\delta\langle\NUE\rangle$ \\
    \midrule
    \mtag{top3}, $s=0.15$ & $-1.13\pm0.17$ & $-0.56$ & $-0.52\pm0.07$ & $+2.5$ \\
    \mtag{top3}, $s=0.075$ & $-0.90\pm0.21$ & $-0.45$ & $-0.41\pm0.08$ & $+1.4$ \\
    \mtag{top5} & $-0.73\pm0.21$ & $-0.37$ & $-0.41\pm0.09$ & $+0.2$ \\
    \mtag{top1} & $-0.84\pm0.21$ & $-0.42$ & $-0.27\pm0.08$ & $+1.1$ \\
    \mtag{m0\_rangeLo} & $-0.40\pm0.21$ & $-0.20$ & $-0.24\pm0.08$ & $+0.8$ \\
    \mtag{cr\_off} & $-0.34\pm0.21$ & $-0.17$ & $+0.24\pm0.08$ & $+6.7$ \\
    \midrule
    \mtag{bfrag\_lo} ($r_b=0.70$) & $+0.03\pm0.21$ & $+0.02$ & $-0.15\pm0.08$ & $-0.2$ \\
    \mtag{bfrag\_hi} ($r_b=1.00$) & $-0.16\pm0.21$ & $-0.08$ & $-0.01\pm0.08$ & $-0.1$ \\
    \midrule
    $\mtmc\pm1\GeV$ (per GeV) & $+0.27\pm0.15$ & $+0.14$ & $+0.68\pm0.06$ & $\approx0$ \\
    \bottomrule
  \end{tabular}
\end{table*}

For all top-specific models and for the reduced CR range, the observed $\mlb$ shift agrees with half the in-cone charged-energy shift within about $1.5\sigma$. For top-specific CR at $s=0.15$ the response proxy changes by about $-1\%$ and accounts for the observed $\mlb$ shift. Through the slope of \cref{sec:theory-mlb}, and with $\langle\mlb\rangle\approx107\GeV$, the mean of the selected $\mlb$ distribution in the 30--160\,GeV fit window (not the unrestricted parton-level mean of 94.1\,GeV), this predicts
\begin{equation}
  \Delta\mtmc \approx \frac{\tfrac12\,(-0.011)\,\langle\mlb\rangle}{0.706} \approx \frac{-0.59\GeV}{0.706} \approx -0.8\GeV,
  \label{eq:dm-estimate}
\end{equation}
close to the template fit shift of $-0.73\pm0.14\GeV$ in \cref{tab:bias}. The charged particles change in a matching way: the cones contain more particles, each with lower $\pT$, and more particles appear outside the jets. This fits a simple picture in which CR stretches the strings from the $b$ quark towards partons elsewhere in the event, so that energy leaves the cone and is shared among more hadrons. We do not derive this pattern in detail.

Two cases do not follow \cref{eq:jes}, and both are informative. Switching CR off raises $\langle\mlb\rangle$, although it lowers the charged energy in the cones, and it adds 6.7 charged particles (15\%) outside the jets. Its mass shift is therefore not a jet-energy effect. Its charged-particle pattern is also very different from that of top-specific CR, which is why the goodness-of-fit test flags it with $\chimin\approx4\times10^4$ (\cref{sec:oof12}). A change of $\mtmc$ moves $\langle\mlb\rangle$ by $0.68\%$ per GeV through the decay kinematics of \cref{eq:mlb-kin} but the in-cone charged energy by only $0.27\%$ per GeV, and leaves the UE unchanged. Mass and CR thus move the observables in different directions, which is the basis of the in-situ method. The two $r_b$ variations confirm the picture of \cref{sec:theory-frag}: they change the in-cone multiplicity and mean $\pT$ in the same way as CR (\cref{fig:plane}) but leave the in-cone charged energy and the UE unchanged within uncertainties.

\section{Simulation setup and event selection}
\label{sec:setup}

The leading-order samples used in \crefrange{sec:setup}{sec:oof12} are generated with \textsc{Pythia}~8.318~\cite{Bierlich:2022pfr} using the default Monash 2013 tune~\cite{Skands:2014pea} (\texttt{Tune:ee = 7}, \texttt{Tune:pp = 14}), at $\sqrts=13.6\TeV$. In \textsc{Pythia}~8.318 this tune switches on the flag \texttt{StringZ:\allowbreak useOldAExtra}, which keeps the pre-8.315 fragmentation handling the tune was fitted with~\cite{Pythia8318doc}, all samples share it. The hard process is leading-order $gg,q\bar q\to\ttbar$, with both $W$ bosons forced to $e\nu$ or $\mu\nu$. Each sample has $7\times10^5$ generated events, giving about $1.04\times10^5$ selected events (acceptance $\approx0.15$). The LO cross section reported by \textsc{Pythia} for $\ttbar$ with both $W$ bosons decaying to $e\nu$ or $\mu\nu$ ($ee$, $\mu\mu$ and $\emu$ combined, before selection) is $26.7\pb$, so a sample corresponds to roughly $26\ifb$. The decay chain and the observable regions are sketched in \cref{fig:diagrams}.

\begin{figure*}[tbp]
  \centering
  \includegraphics[width=0.97\textwidth]{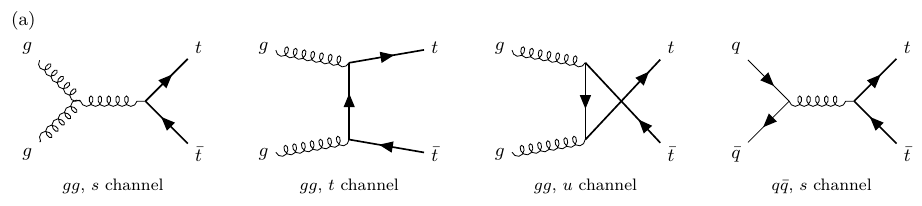}\\[8pt]
  \includegraphics[width=0.80\textwidth]{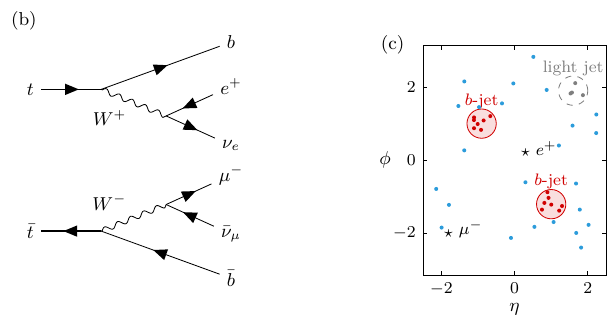}\\[6pt]
  \includegraphics[width=\textwidth]{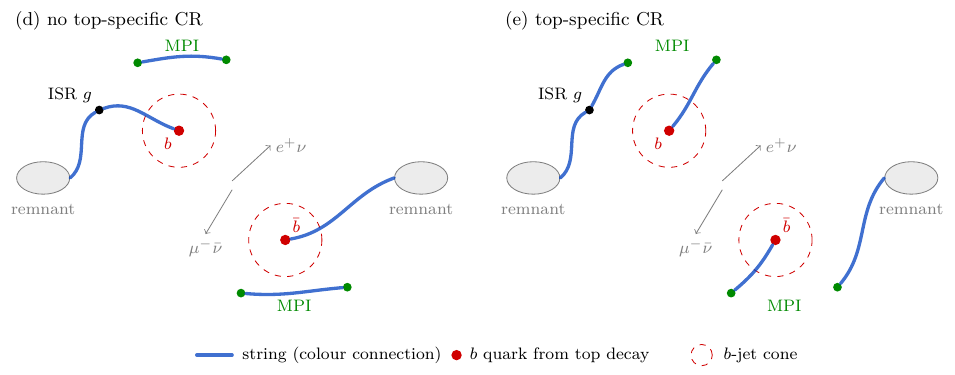}
  \caption{(a) The four leading-order diagrams for top-pair production: $gg\to\ttbar$ through the $s$, $t$ and $u$ channels, and $q\bar q\to\ttbar$. Gluon fusion gives most of the cross section at the LHC, both channels are included in all samples, and the NLO samples also contain the real and virtual corrections, including $qg$-initiated production. (b) The top-quark decays selected in the $\emu$ channel. (c) Schematic $\eta$--$\phi$ view of the observables: charged particles ($\pT>0.5\GeV$, $|\eta|<2.5$) inside the two $b$-jet cones ($\Delta R<0.4$, red) and outside all jets (blue). (d, e) Schematic colour flow of the same event. Without top-specific CR (d), the strings attached to the $b$ quarks run to the partons they are colour-connected to in the hard process and the shower, here an initial-state gluon and a beam remnant, while the MPI partons form their own strings. With top-specific CR (e), the strings attached to the top-decay partons can reconnect with partons from the rest of the event, here MPI partons, which changes the hadrons produced in and around the $b$-jet cones. The leptons and neutrinos are colour neutral and carry no strings. Panels (c)--(e) are schematic, not simulated, ISR: initial-state radiation.}
  \label{fig:diagrams}
\end{figure*}

\paragraph{Selection} Events must contain exactly one electron and one muon of opposite charge, each with $\pT>25\GeV$ and $|\eta|<2.5$. Leptons are taken as bare final-state leptons. Jets are clustered with the anti-$k_t$ algorithm with $R=0.4$~\cite{Cacciari:2008gp}, from all visible final-state particles except the two leptons, and must have $\pT>25\GeV$ and $|\eta|<2.5$. A jet is $b$-tagged if a weakly decaying $B$ hadron lies within $\Delta R<0.4$. Exactly two $b$-jets are required.

\paragraph{Charged particles} We use final-state charged particles with $\pT>0.5\GeV$ and $|\eta|<2.5$, excluding the two leptons. A particle is \emph{in a $b$-jet} if it lies within $\Delta R<0.4$ of either $b$-jet axis. It is \emph{underlying event} (UE) if it lies outside $\Delta R=0.4$ of every selected jet, as in the ATLAS CR-sensitive measurement~\cite{ATLAS:2022zbu}. Unstable hadrons are decayed with the \textsc{Pythia} defaults, so charged decay products of $K^0_S$, $\Lambda$ and other strange hadrons are included. This differs from the $\tau>30$\,ps convention of the ATLAS measurement, which is applied in the comparison with data (\cref{sec:rivet-tau}).

\paragraph{Models} The nominal model is MPI-based CR (mode~0) with late resonance decays. The variations fall into three groups, listed in \cref{tab:models}, exact settings are given in \cref{app:generator}.

\begin{table*}[tbp]
  \centering
  \caption{Generator variations and the model tags used in all tables. All differ from the nominal (MPI-based CR, late resonance decays, Monash~2013) only in the settings listed.}
  \label{tab:models}
  \small
  \begin{tabular}{lll}
    \toprule
    Group & Tag & Setting changed from nominal \\
    \midrule
    \multirow{2}{*}{Top-specific CR} & \mtag{top1}--\mtag{top5} & TopReconUserHooks mode 1--5~\cite{Argyropoulos:2014zoa}, strength 0.075 (mode 5: 1.0) \\
     & \mtag{top3s<s>} & mode 3 with continuous strength $s\in[0,0.15]$ \\
    \midrule
    \multirow{5}{*}{Generic CR} & \mtag{cr\_off} & CR switched off \\
     & \mtag{m0\_rangeLo/Hi} & CR range 1.2 / 2.4 (default 1.8) \\
     & \mtag{m0\_early} & early resonance decays \\
     & \mtag{m1} & QCD-based CR~\cite{Christiansen:2015yqa}, not retuned \\
     & \mtag{m2} & gluon-move CR \\
    \midrule
    \multirow{4}{*}{Non-CR} & \mtag{m0\_pt215} & MPI $p_{\mathrm{T}0}^{\mathrm{ref}}=2.15\GeV$ (default 2.28) \\
     & \mtag{bfrag\_lo/hi} & Bowler $r_b=0.70$ / $1.00$ (default 0.855) \\
     & \mtag{bfrag\_pet} & Peterson $b$ fragmentation \\
     & \mtag{frag\_aLund} & Lund $a=0.78$ (default 0.68) \\
    \bottomrule
  \end{tabular}
\end{table*}

The nominal model is generated at $\mtmc=171.5$, $172.0$, $172.5$, $173.0$ and $173.5\GeV$. Variations are generated at $172.5\GeV$ unless stated. The nominal sample at $172.5\GeV$ is generated with two independent random seeds, and five further repeated samples of selected configurations measure the seed-to-seed scatter directly.

\paragraph{Tracking} A 1\% loss of charged-track efficiency is simulated in a simple way, by removing particles from the counts. In each event, each particle counted in $\Ninb$ and $\NUE$ is kept with probability 0.99, $\SUE$ is scaled by the fraction of particles kept outside the jets, and the tail fractions are recomputed from the reduced $\Ninb$. The mean $\ptinb$ and its quantiles are left unchanged, a track-by-track treatment is left for a detector-level study. It enters the inference as a nuisance (\cref{sec:inference}).

\section{The \texorpdfstring{$m_{\ell b}$}{mlb} estimator and the CR-induced bias}
\label{sec:mlb}

Top-specific CR moves the template-fit mass by $-0.43$ to $-0.73\GeV$, larger than the total uncertainty of the current ATLAS+CMS combination. Generic CR variants move it by $-0.35$ to $+0.28\GeV$.

\paragraph{Estimator} Each $b$-jet is paired with a lepton using the minimax rule, which picks the pairing whose larger invariant mass is smaller:
\begin{equation}
  \begin{split}
  \mlbmm = \min\Big[&\max\big(m_{\ell^+ b_1}, m_{\ell^- b_2}\big),\\
  &\max\big(m_{\ell^+ b_2}, m_{\ell^- b_1}\big)\Big].
  \end{split}
  \label{eq:minimax}
\end{equation}
For a correct pairing at parton level, $\mlb$ is bounded by the endpoint $\sqrt{\mt^2-\mW^2}\approx152.6\GeV$ at $\mt=172.5\GeV$ (\cref{fig:mlb}). The fit uses $30<\mlb<160\GeV$.

\begin{figure}[tbp]
  \centering
  \includegraphics[width=\columnwidth]{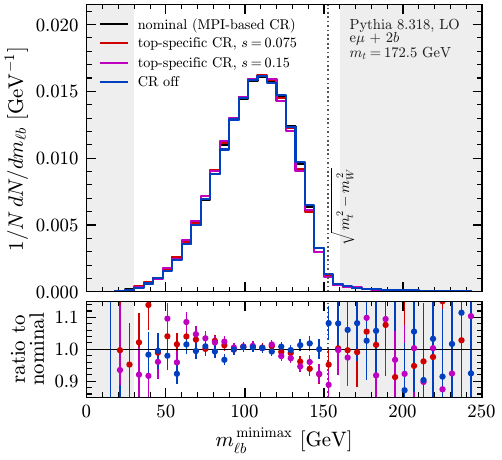}
  \caption{Minimax $\mlb$ for the nominal sample, top-specific reconnection (TopRecon mode 3, $s=0.075$ and $0.15$) and CR off, normalised, lower panel: ratio to nominal. Dotted line: kinematic endpoint $\sqrt{\mt^2-\mW^2}$. Shaded: outside the 30--160\,GeV fit window.}
  \label{fig:mlb}
\end{figure}

\paragraph{Template fit} Normalised $\mlb$ histograms with $5\GeV$ bins are built from the four nominal samples at $171.5$, $172.0$, $173.0$ and $173.5\GeV$. The $172.5\GeV$ reference sample is excluded so that it is never fitted against itself. In each bin, the template fraction is fitted as a straight line in $\mt$ by weighted least squares. The fitted mass minimises a $\chi^2$ whose per-bin variance includes both the sample count and the template uncertainty, scanned on a $10\MeV$ grid. The mean $\mlb$ in the window rises by $0.706\pm0.046\GeV$ per GeV of $\mt$ across the five nominal masses (\cref{fig:calib}), with an independent replica at $172.5\GeV$ consistent with the fit.

\begin{figure}[tbp]
  \centering
  \includegraphics[width=\columnwidth]{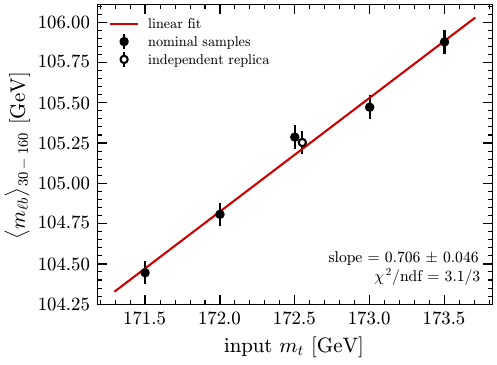}
  \caption{Mean $\mlb$ in the fit window versus input $\mt$ for five nominal samples, with a linear fit, open marker: independent replica at $172.5\GeV$.}
  \label{fig:calib}
\end{figure}

\paragraph{Bias} \Cref{tab:bias} lists the shift $\dmtmc$ of the fitted mass for each variation, relative to the mean of the two nominal seeds at $172.5\GeV$ (\cref{fig:massbias}). Uncertainties come from bootstrap resampling with shared template fluctuations. These are shifts of the generator mass parameter, not experimental systematic uncertainties. The uncertainty of each shift is 0.11--0.14\,GeV, set by the statistics of the compared samples. Shifts below about $0.25\GeV$ (gluon move, QCD-based CR, MPI, $r_b=1.00$, Peterson and Lund $a$) are individually consistent with zero.

\begin{table*}[tbp]
  \centering
  \caption{Shift of the $\mlb$ template-fit mass relative to the mean of two nominal seeds at $\mtmc=172.5\GeV$. Model tags as in \cref{tab:models}. Uncertainties: bootstrap standard deviation with shared template fluctuations.}
  \label{tab:bias}
  \small
  \begin{tabular}{llr}
    \toprule
    Group & Model & $\dmtmc$ [GeV] \\
    \midrule
    \multirow{4}{*}{Top-specific CR} & \mtag{top3}, $s=0.15$ & $-0.73\pm0.14$ \\
     & \mtag{top3}, $s=0.075$ & $-0.55\pm0.11$ \\
     & \mtag{top5} & $-0.51\pm0.12$ \\
     & \mtag{top1} & $-0.43\pm0.12$ \\
    \midrule
    \multirow{4}{*}{Generic CR} & \mtag{m0\_rangeLo} & $-0.35\pm0.11$ \\
     & \mtag{m2} & $-0.01\pm0.12$ \\
     & \mtag{m1} (untuned) & $+0.00\pm0.12$ \\
     & \mtag{cr\_off} & $+0.28\pm0.12$ \\
    \midrule
    \multirow{5}{*}{Non-CR} & \mtag{bfrag\_lo} ($r_b=0.70$) & $-0.29\pm0.11$ \\
     & \mtag{bfrag\_hi} ($r_b=1.00$) & $-0.10\pm0.11$ \\
     & \mtag{frag\_aLund} & $-0.05\pm0.11$ \\
     & \mtag{bfrag\_pet} & $+0.03\pm0.11$ \\
     & \mtag{m0\_pt215} & $+0.09\pm0.11$ \\
    \bottomrule
  \end{tabular}
\end{table*}

\begin{figure}[tbp]
  \centering
  \includegraphics[width=\columnwidth]{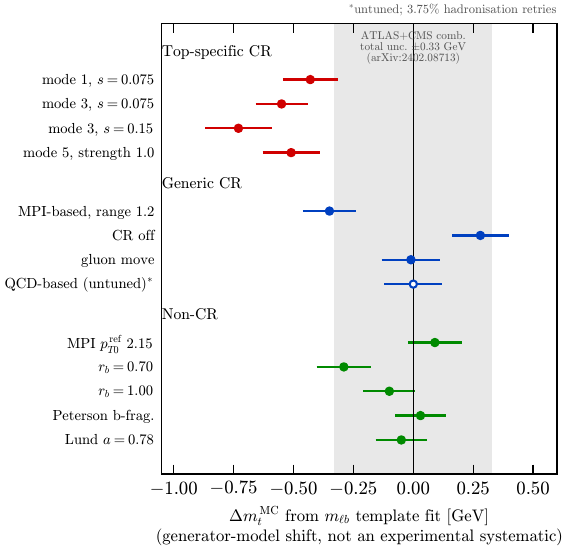}
  \caption{Shift of the $\mlb$ template-fit mass, $\dmtmc$, for each generator variation relative to the mean of two nominal seeds, errors from bootstrap with shared template fluctuations. These are generator-model shifts in $\mtmc$, not experimental systematic uncertainties, the band shows the ATLAS+CMS combined total uncertainty~\cite{ATLAS:2024dxp} for scale only. Open marker: QCD-based CR without a dedicated tune (3.75\% hadronisation retries).}
  \label{fig:massbias}
\end{figure}

Two features matter for what follows. Top-specific CR always lowers the fitted mass, and the shift grows with the strength $s$. This pattern, large negative shifts from reconnections of the top-decay products and smaller shifts of either sign from generic models, agrees with the semileptonic study of Ref.~\cite{Argyropoulos:2014zoa}. The softer Bowler setting $r_b=0.70$ gives a shift comparable to the weaker CR models, so any CR-sensitive observable must also separate CR from $b$-fragmentation.

\section{A \texorpdfstring{$b$}{b}-jet charged-particle fingerprint}
\label{sec:fingerprint}

Charged particles inside the $b$-jets respond strongly to CR and barely to $\mt$. Their means alone confuse CR with $b$-fragmentation, but adding distribution shapes separates all five CR model pairs we tested. One fragmentation variant can still mimic a CR model.

\paragraph{Observables} Per event we record the charged multiplicity $\Ninb$ inside the two $b$-jet cones and their mean transverse momentum $\ptinb$, plus the multiplicity $\NUE$ and scalar $\pT$ sum $\SUE$ outside all jets. From these the flow vector has nine entries:
\begin{enumerate}
  \item the means of $\Ninb$, $\ptinb$, $\NUE$ and $\SUE$,
  \item the tail fractions $f(\Ninb\le10)$ and $f(\Ninb\ge26)$,
  \item the 10\%, 50\% and 90\% quantiles of $\ptinb$.
\end{enumerate}

\paragraph{Mass versus CR response} Across the five nominal masses, the flow means change by at most $0.019\pm0.005\GeV$ per GeV of $\mt$ ($\ptinb$), the multiplicities are consistent with no dependence. Top-specific CR instead raises $\langle\Ninb\rangle$ by 0.13--0.25 and lowers $\langle\ptinb\rangle$ by 0.08--0.12\,GeV (\cref{fig:fingerprint,fig:plane}). The flow therefore measures CR without absorbing the mass.

\begin{figure*}[tbp]
  \centering
  \includegraphics[width=\textwidth]{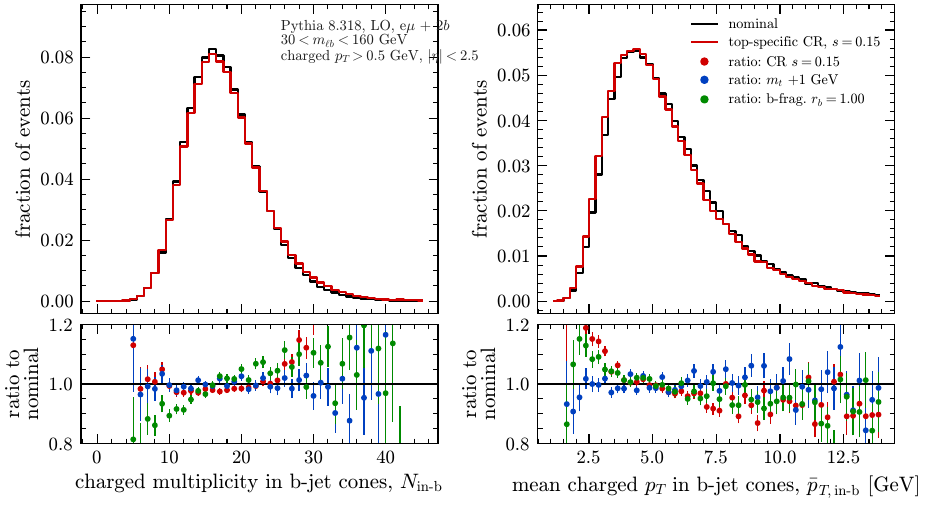}
  \caption{Charged multiplicity and mean charged $\pT$ inside the $b$-jet cones for the nominal sample and top-specific CR ($s=0.15$), lower panels: ratios to nominal for CR, $\mt+1\GeV$ and Bowler $r_b=1.00$.}
  \label{fig:fingerprint}
\end{figure*}

\paragraph{Degeneracy at the level of means} When only the two mean values in the cones are used (\cref{fig:plane}), top-specific CR, CR off, $r_b=1.00$ and Peterson fragmentation all shift them in nearly the same direction, so a fit using only these means cannot tell them apart. For example, with $\mt$, MPI and tracking as free parameters, the $\chi^2$ between \mtag{top1} and the gluon-move model, using the charged particles only, is 0.06 for one degree of freedom, although their mass biases differ by about $0.4\GeV$.

\begin{figure}[tbp]
  \centering
  \includegraphics[width=\columnwidth]{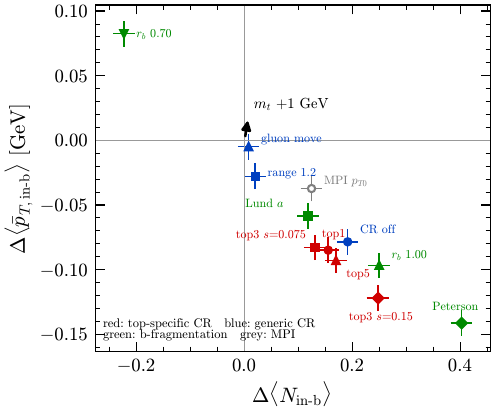}
  \caption{Shifts of the mean in-cone multiplicity and mean charged $\pT$ for all variations, arrow: response to $\mt+1\GeV$. At the level of means, fragmentation and CR off overlap with top-specific CR.}
  \label{fig:plane}
\end{figure}

\paragraph{Shapes break it} Adding the tail fractions and $\ptinb$ quantiles raises the pairwise $\chi^2$ to roughly 65--300 for the five pairs in \cref{fig:shapes}, well above the nominal 95\% quantiles (about 12.6 for six degrees of freedom). The separation survives profiling $r_b$ as an extra nuisance. CR changes how particles are shared within the cone, which the tested Bowler variations cannot reproduce once the full distribution is used, this does not hold for every fragmentation model (\cref{fig:impostors}).

\begin{figure}[tbp]
  \centering
  \includegraphics[width=\columnwidth]{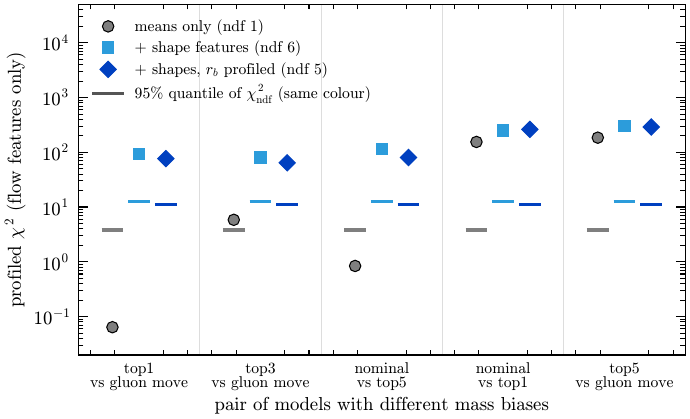}
  \caption{Flow-only profiled $\chi^2$ between model pairs with different mass biases, using means only, adding shape features, and additionally profiling $r_b$ ($\mt$, MPI and tracking profiled throughout). Black ticks: nominal $\chi^2$ 95\% quantiles, an uncalibrated (Asimov-like) reference.}
  \label{fig:shapes}
\end{figure}

\paragraph{Impostor tests} We fit two fragmentation variants that are not part of any nuisance, Peterson $b$-fragmentation and Lund $a=0.78$, to each CR hypothesis with $\mt$, MPI, tracking and $r_b$ profiled (\cref{fig:impostors}). Peterson is rejected by every hypothesis, closest to nominal ($\chi^2\approx21$). Lund $a=0.78$ is rejected by all hypotheses except gluon move, where $\chi^2$ falls just below the nominal 95\% quantile. Fragmentation changes outside the nuisance model can therefore mimic a CR model, and the goodness-of-fit test of \cref{sec:oof12} is not a guarantee against this.

\begin{figure}[tbp]
  \centering
  \includegraphics[width=\columnwidth]{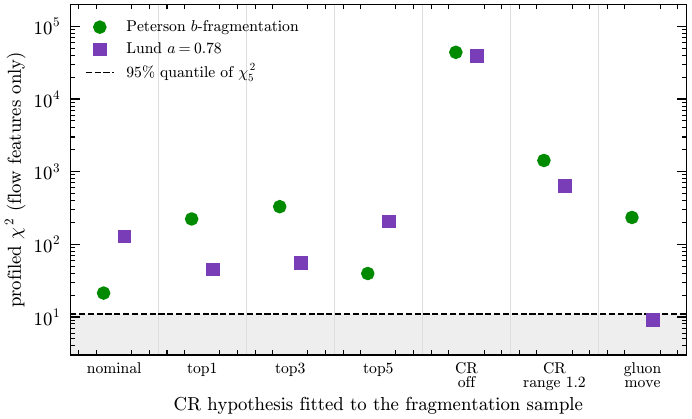}
  \caption{Profiled $\chi^2$ of Peterson and Lund-$a$ fragmentation samples to each CR hypothesis ($\mt$, MPI, tracking, $r_b$ profiled), dashed: nominal $\chi^2_5$ 95\% quantile (uncalibrated reference).}
  \label{fig:impostors}
\end{figure}

The $\chi^2$ references here are nominal quantiles from Asimov-like comparisons, not calibrated thresholds, the calibrated threshold used for decisions is defined in \cref{sec:oof12}.

\section{Inference within a continuous CR family}
\label{sec:inference}

We fit the fitted mass and the nine flow features jointly, with the top-specific CR strength $s$ as a free parameter, so that the flow fixes $s$ and the mass is corrected for it.

\paragraph{Training family} The family is TopRecon mode 3 with strength $s=0$, $0.0375$, $0.075$, $0.1125$ and $0.15$, where $s=0$ is the nominal model. Each strength has two independent seeds at $\mt=172.5\GeV$, and each training point is the mean of the two. Over this range the fitted mass and the flow features respond close to linearly in $s$ (\cref{fig:strength}).

\begin{figure*}[tbp]
  \centering
  \includegraphics[width=\textwidth]{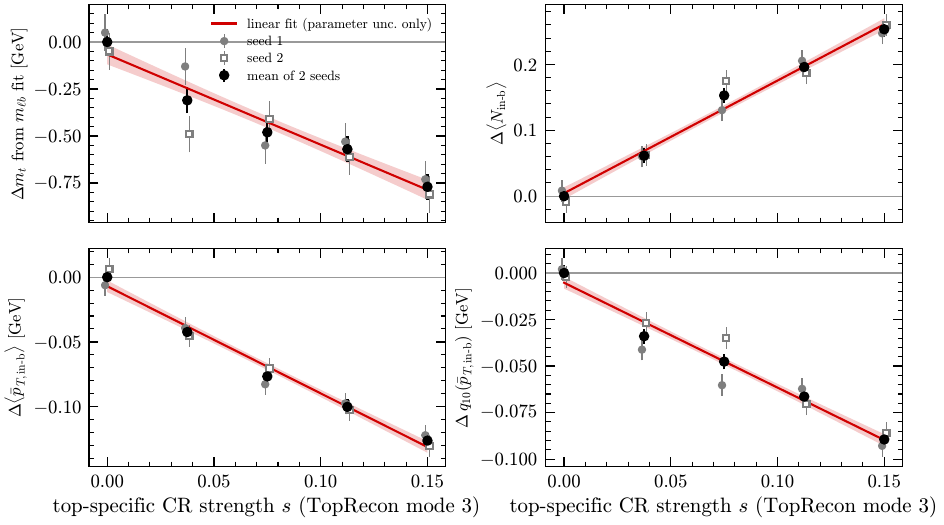}
  \caption{Response of the fitted mass and three flow features to the top-specific CR strength $s$ (two independent seeds per strength and their mean), line and band: linear fit and its parameter uncertainty only.}
  \label{fig:strength}
\end{figure*}

\paragraph{Model} For a sample, the ten-component observable is $y=(m_{\mathrm{fit}},\text{flow})-z_{\mathrm{ref}}$, where $z_{\mathrm{ref}}$ is the nominal reference. Its expectation is
\begin{equation}
  \mathbb{E}[y] = \Delta\mt\, g + a + b\, s + \sum_j \nu_j\, n_j ,
  \label{eq:model}
\end{equation}
where $\Delta\mt=\mtmc-172.5\GeV$, $g$ is the response to a $1\GeV$ change of $\mt$ from the five nominal mass points, $a+bs$ is the least-squares line through the training points, and $n_j$ are nuisance directions with constrained parameters $\nu_j\sim\mathcal{N}(0,1)$. The three nuisances are MPI ($p_{\mathrm{T}0}^{\mathrm{ref}}=2.15\GeV$ minus nominal), tracking (the count-level 1\% proxy of \cref{sec:setup}, averaged over 20 thinnings) and $b$-fragmentation (half of the $r_b=1.00$ minus $r_b=0.70$ response).

\paragraph{Covariance} $V_1$ is the covariance of $y$ for one sample, from 300 bootstrap replicas of the reference. It describes the statistical fluctuation of the tested sample only, the $\mlb$ templates, the training points and $z_{\mathrm{ref}}$ are single fixed realisations. The residual covariance adds the uncertainty of the fitted line at $s$:
\begin{equation}
  C(s) = \Big(1+\frac{\kappa(s)}{2}\Big)V_1, \qquad \kappa(s) = x_s^{\top}\big(X^{\top}X\big)^{-1}x_s ,
  \label{eq:cov}
\end{equation}
with $x_s=(s,1)$, $X$ the design matrix of the training strengths, and the factor 2 for the two seeds averaged at each training point.

\paragraph{Likelihood} At each $s$ on a 61-point grid over $[0,0.15]$, $\Delta\mt$ and $\nu$ are fitted by generalised least squares with the $\mathcal{N}(0,1)$ penalty on $\nu$, giving
\begin{equation}
  -2\ln L(s) = \chimin(s) + \ln\det C(s) .
  \label{eq:lik}
\end{equation}
The profile likelihood in $\Delta\mt$ is the minimum over $s$, and the 68\% and 95\% intervals are where it rises by 1 and 3.84. The best-fit $\chimin$ serves as the goodness-of-fit statistic.

\paragraph{Pre-specified variants} Four methods were fixed before any blind test (\cref{tab:methods}), only the first is used for decisions.

\begin{table*}[tbp]
  \centering
  \caption{Inference methods fixed before the blind tests. Only \mtag{primary} is used for decisions, \mtag{mass\_only} is the $\mlb$ fit alone, with an interval widened to cover the CR family. $V_1$: bootstrap covariance of the mass and flow vector for one sample (\cref{sec:inference}).}
  \label{tab:methods}
  \small
  \begin{tabular}{llll}
    \toprule
    Method & Response in $s$ & Covariance & Observables \\
    \midrule
    \mtag{primary} & linear & $V_1$ (bootstrap) & mass + 9 flow \\
    \mtag{alt\_scaled} & linear & $V_1$ inflated by seed-pair scatter & mass + 9 flow \\
    \mtag{alt\_quad} & quadratic & $V_1$ & mass + 9 flow \\
    \mtag{mass\_only} & linear & $V_1$ & mass only \\
    \bottomrule
  \end{tabular}
\end{table*}

\paragraph{Corrections made before freezing} An external review of earlier versions found four errors, which we fixed: a covariance that was too large by a factor $2(1+\kappa)$ on $V_1$, a missing log-determinant term, template fluctuations that were not shared between pseudo-experiments, and the reference sample included in its own templates. An earlier claimed gain of $3.8\times$ in precision came from the overcounted covariance and is withdrawn. The frozen version (v3) contains all fixes, its hashes are recorded in the analysis log (\cref{app:blinding}).

\section{Blind validation}
\label{sec:blind}

The estimator of \cref{sec:inference} was tested in three blind tests at leading order and particle level, run in sequence. Each followed the same protocol (\cref{app:blinding}). The method was fixed and its files hashed. The samples were generated and anonymised, and the key was kept private. The predictions were hashed before the key was opened, and only then were they scored.
 The first test (\cref{sec:val24}) measures bias and coverage inside the trained CR family, the second (\cref{sec:oof12}) whether the goodness-of-fit test detects models outside it, and the third (\cref{sec:cms7}) uses independently tuned CMS models. We refer to the tests by the names used in the analysis log and code: val24, oof12 and cms7.

\subsection{Within the trained model family (val24)}
\label{sec:val24}

On 24 blinded samples the frozen primary method covered the true mass at close to nominal rates, with no measurable bias and about half the root-mean-square error of the mass-only fit.

\paragraph{Design} The 24 samples use TopRecon mode 3 at three strengths not in the training set ($s=0.05$, $0.09$, $0.13$, eight each) and $\mt=172.0$, $172.5$ or $173.0\GeV$, with new seeds. Files were anonymised and the key was held back by the author. The method was frozen and its files hashed, predictions for all four methods were written and their SHA-256 hash posted before the key was opened (\cref{app:blinding}).

\paragraph{Results} \Cref{tab:val24} gives the scores. For the primary method the 68\% intervals covered 16 of 24 samples (Wilson 95\% interval 0.47--0.82) and the 95\% intervals covered all 24. The mean of fitted minus true mass was $-0.017\pm0.020\GeV$, the pull width 0.94, and no sample exceeded the goodness-of-fit threshold. The recovered strengths average about 0.058, 0.096 and 0.130 for the three true values (\cref{fig:val24}).

\begin{table*}[tbp]
  \centering
  \caption{Scores of the four pre-specified methods (defined in \cref{tab:methods}) on the 24 blinded in-family samples. Errors are fitted minus true mass. Mean bias: their mean, spread: their sample standard deviation, RMSE: their root mean square, which includes the bias ($\mathrm{RMSE}^2\approx\text{bias}^2+\text{spread}^2$). Coverage: fraction of samples whose interval contains the true mass. Pull width: standard deviation of the pulls (error divided by its reported uncertainty).}
  \label{tab:val24}
  \small
  \begin{tabular}{lccccccc}
    \toprule
    Method & \multicolumn{2}{c}{Coverage} & Mean bias & Spread & RMSE & Median 68\% & Pull \\
     & 68\% & 95\% & [GeV] & [GeV] & [GeV] & half-width [GeV] & width \\
    \midrule
    \mtag{primary} & 16/24 & 24/24 & $-0.017\pm0.020$ & 0.096 & 0.096 & 0.101 & 0.94 \\
    \mtag{alt\_scaled} & 20/24 & 24/24 & $-0.016\pm0.020$ & 0.097 & 0.096 & 0.140 & 0.70 \\
    \mtag{alt\_quad} & 18/24 & 24/24 & $-0.009\pm0.020$ & 0.097 & 0.095 & 0.103 & 0.92 \\
    \mtag{mass\_only} & 24/24 & 24/24 & $-0.110\pm0.030$ & 0.146 & 0.181 & 0.495 & 0.30 \\
    \bottomrule
  \end{tabular}
\end{table*}

\begin{figure*}[tbp]
  \centering
  \includegraphics[width=\textwidth]{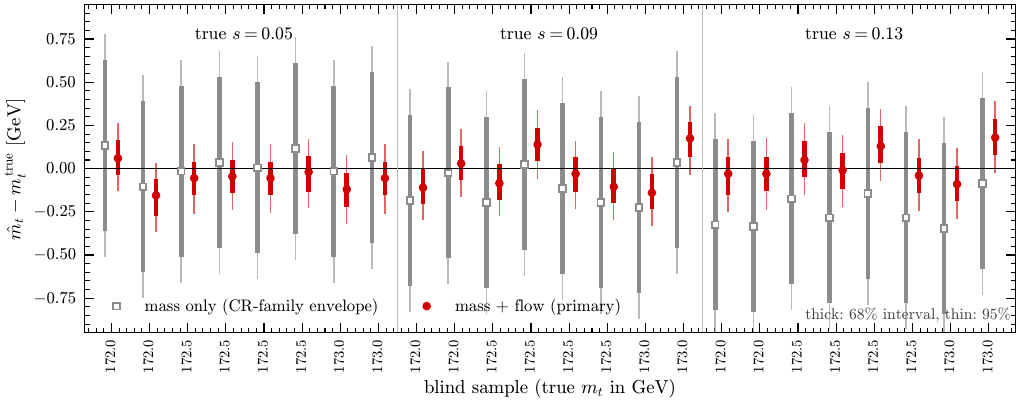}
  \caption{Frozen, pre-registered validation on 24 independently generated samples at three untrained strengths and three masses, evaluated against one common frozen training set. Top: fitted-minus-true $\mt$ with 68\% intervals for mass-only (CR-family envelope) and mass+flow (primary). Bottom: pulls of the primary method, empirical coverage with Wilson 95\% bands, and median 68\% half-widths. Primary: 16/24 (68\%), 24/24 (95\%), bias $-0.017\pm0.020\GeV$. The mass-only envelope is deliberately conservative, the width ratio is not a matched-coverage comparison. Coverage is conditional on one frozen training set.}
  \label{fig:val24}
\end{figure*}

\paragraph{Comparison with mass only} Without the charged particles, the mass-only method cannot determine $s$, so its interval must cover the whole CR family, its bias grows from $+0.03\GeV$ at $s=0.05$ to $-0.25\GeV$ at $s=0.13$. Its root-mean-square error is $0.18\GeV$ against $0.096\GeV$ for the primary method. The ratio of interval half-widths, $0.495/0.101\approx4.9$, is not a fair measure of the gain, because the mass-only interval is too wide (it covers 24/24 at 68\%, pull width 0.30). The gain over a mass-only method with equally good coverage has not been measured and would need fresh samples.

\paragraph{Caveats} The 68\% coverage is at the low edge of nominal, and lowest at $s=0.09$ (4/8) and $\mt=173.0\GeV$ (2/6). The coverage refers to one fixed set of inputs: one set of $\mlb$ templates, one sample per training point, one reference, and a covariance bootstrapped from one nominal sample. It is not an average over independently regenerated inputs. Because all 24 results share these inputs, they are correlated, and the number of independent checks is smaller than 24. All samples are in-family by construction, \cref{sec:oof12} tests what happens when they are not.

\subsection{Outside the trained model family (oof12)}
\label{sec:oof12}

The pre-registered criteria for claim B were met: both CR-off samples were flagged, there were no false alarms and no critical failures. Detection is model-specific, however, and four of the six other alternative models passed unflagged.

\paragraph{Pre-registration} Before any oof12 event existed we fixed the composition, the decision rule and the interpretation, and hashed every file. The 12 samples are 2 CR off, 6 other CR models (\mtag{top1}, \mtag{top2}, \mtag{top4}, \mtag{m0\_early}, \mtag{m0\_rangeHi}, \mtag{m2}) and 4 in-family controls with $s$ drawn from $\mathcal{U}[0.02,0.14]$. Masses were drawn from $\{172.0,172.5,173.0\}\GeV$ with fresh seeds. Of the six other models, \mtag{top1} and \mtag{m2} had been analysed in the pilot, \mtag{top2}, \mtag{top4}, \mtag{m0\_early} and \mtag{m0\_rangeHi} had not.

\paragraph{Decision rule} A sample is flagged if $\chimin$ of the primary fit exceeds 36.67. This cutoff was computed from training samples only, before val24 existed. Each interior strength ($s=0.0375$, $0.075$, $0.1125$) was removed from the training set in turn, with both its seeds. Each held-out seed was then fitted 150 times, with Poisson-fluctuated templates and a bootstrap of the held-out events. The cutoff is the 99th percentile of the resulting 900 values of $\chimin$. These come from only six underlying samples, so the 1\% false-alarm rate is a training-derived target, not a demonstrated rate. Unflagged samples report the primary interval. Flagged samples report a fallback built from the mass-only fit $m$ and a CR envelope $b\in[-0.80,+0.30]\GeV$:
\begin{equation}
  \begin{gathered}
  \Delta\mt \in \big[\, m - b_{\max} - k\sigma,\ \ m - b_{\min} + k\sigma \,\big],\\
  k = 1\ (68\%),\ 1.96\ (95\%),
  \end{gathered}
  \label{eq:fallback}
\end{equation}
where $m=m_{\mathrm{fit}}-m_{\mathrm{fit,ref}}$ is the mass-only template-fit result relative to the nominal reference, in the same convention as $\Delta\mt=\mtmc-172.5\GeV$, and $\sigma$ is its statistical error from the frozen covariance. Claim B required four things: no critical failure (an unflagged sample from outside the family whose 95\% interval misses the true mass), both CR-off samples flagged, at most one false alarm among the controls, and every flagged out-of-family sample covered by its 95\% fallback interval.

\paragraph{Results} \Cref{tab:oof12} lists every sample and \cref{fig:oof12} shows the reported intervals. The reported intervals covered 6/12 at 68\% and 11/12 at 95\%, the narrow primary interval alone would have covered 4/12 and 8/12. Without the fallback, the CR-off samples would have been reported $1.42\GeV$ low with intervals of about $\pm0.1\GeV$.

\begin{table*}[tbp]
  \centering
  \caption{Per-sample results of the out-of-family test. ``Primary error'' is the narrow primary fit minus the true mass, the last column refers to the reported (hybrid) interval. Model tags as in \cref{tab:models}, $\chimin$ is the goodness-of-fit statistic of the flow features (\cref{sec:inference}).}
  \label{tab:oof12}
  \small
  \setlength{\tabcolsep}{4pt}
  \begin{tabular}{lllrcrc}
    \toprule
    Sample & True model & $\mt$ [GeV] & $\chimin$ & Flagged & Primary error [GeV] & Covers 68\% / 95\% \\
    \midrule
    oof\_04 & \mtag{cr\_off} & 172.0 & 42\,271 & yes & $-1.420$ & no / yes \\
    oof\_12 & \mtag{cr\_off} & 173.0 & 42\,064 & yes & $-1.425$ & no / yes \\
    oof\_03 & \mtag{m0\_rangeHi} & 172.0 & 776 & yes & $+0.335$ & yes / yes \\
    oof\_05 & \mtag{m2} & 172.0 & 43.9 & yes & $+0.110$ & yes / yes \\
    oof\_02 & \mtag{top4} & 173.0 & 25.9 & no & $-0.115$ & no / yes \\
    oof\_07 & \mtag{m0\_early} & 172.0 & 24.8 & no & $+0.085$ & yes / yes \\
    oof\_06 & \mtag{top2} & 172.0 & 24.4 & no & $-0.135$ & no / yes \\
    oof\_08 & \mtag{top1} & 173.0 & 6.5 & no & $+0.065$ & yes / yes \\
    \midrule
    oof\_10 & \mtag{top3}, $s=0.0423$ & 172.0 & 13.7 & no & $-0.135$ & no / yes \\
    oof\_11 & \mtag{top3}, $s=0.134$ & 172.5 & 11.8 & no & $+0.215$ & no / no \\
    oof\_09 & \mtag{top3}, $s=0.0855$ & 172.5 & 3.4 & no & $+0.070$ & yes / yes \\
    oof\_01 & \mtag{top3}, $s=0.0541$ & 172.0 & 3.1 & no & $+0.045$ & yes / yes \\
    \bottomrule
  \end{tabular}
\end{table*}

\begin{figure}[tbp]
  \centering
  \includegraphics[width=\columnwidth]{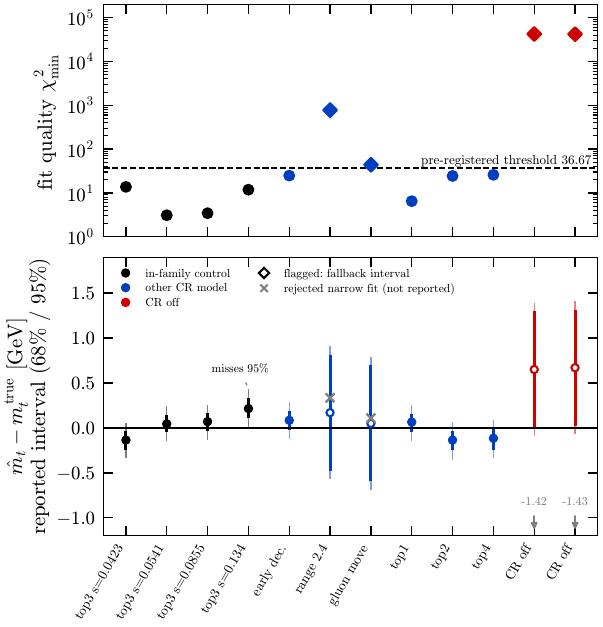}
  \caption{Pre-registered out-of-family blind test (12 samples). Top: fit quality $\chimin$ with the pre-registered threshold (36.67). Bottom: reported interval per sample (68\% thick, 95\% thin). For flagged samples (open markers) the reported interval is the two-sided fallback envelope, the rejected narrow fit is shown only for reference (grey, arrows mark off-scale values, $-1.42\GeV$ for CR off). Pre-registered verdict: 0 critical failures, CR off flagged 2/2, 0/4 false alarms. Reported intervals cover 6/12 (68\%) and 11/12 (95\%), the narrow primary fit alone would cover 4/12 and 8/12. The in-family control at $s=0.134$ misses at 95\%. Both CR-off samples have mass-only shifts ($+0.40$ to $+0.42\GeV$) beyond the pre-registered $+0.30\GeV$ envelope edge and miss at 68\%. \mtag{top1}, \mtag{top2}, \mtag{top4} and early resonance decays were not flagged, their narrow intervals covered at 95\%.}
  \label{fig:oof12}
\end{figure}

\paragraph{What it does not show} By the pre-registered definition, the unflagged \mtag{top1}, \mtag{top2}, \mtag{top4} and \mtag{m0\_early} samples are detection failures, even though their narrow intervals covered at 95\%. Their $\chimin$ values (6.5--25.9) sit well below the threshold. The only 95\% miss was the in-family control at $s=0.134$, an edge case of the linear model. Both CR-off samples have mass-only shifts of $+0.40$ to $+0.42\GeV$, beyond the $+0.30\GeV$ envelope edge set from pilot statistics, they were covered at 95\% only through the statistical term, and missed at 68\%.

\paragraph{Simple comparators} A simple Gaussian interval built from the full pilot ensemble and fixed in advance ($G_{\mathrm{ens}}$: mass-only fit, CR shift uniform on $[-0.80,+0.30]\GeV$, $\sigma_{\mathrm{CR}}=0.318\GeV$) covered 10/12 at 68\% and 11/12 at 95\%. This matches our method at 95\%, but with much wider intervals for unflagged samples (68\% half-width at least $0.32\GeV$ against about $0.10\GeV$). The family-only Gaussian $G_{\mathrm{fam}}$ covered 8/12 and 10/12. The hybrid's advantage is precision when the model is in family, not better coverage when it is not.

With 12 samples this is a stress test, not a measurement of detection efficiency. The post-hoc observation that a threshold near 20 would also have flagged \mtag{top2}, \mtag{top4} and \mtag{m0\_early} is not a result, it would need a new pre-registered test on fresh samples.

\subsection{Complete CMS tunes (cms7)}
\label{sec:cms7}

The frozen method flagged all six samples generated with CMS tunes, but its fallback intervals covered the true mass in only one of them. Under the pre-registered rule this is a failure, and we report it as a limitation of the uncertainty model. The cause is that the CMS tunes shift the $\mlb$ fit, calibrated with Monash templates, by about $+0.6\GeV$, more than the fallback envelope allows.

\paragraph{Design} The test uses three independently tuned CMS configurations: CP5~\cite{CMS:2019csb}, and its two colour-reconnection variants CP5-CR1 (QCD-based CR, which forms junctions~\cite{Christiansen:2015yqa,Bierlich:2015rha}) and CP5-CR2 (gluon move)~\cite{CMS:2022awf}. They differ from the training family in the parton distribution functions (PDF, NNPDF3.1 NNLO~\cite{NNPDF:2017mvq}, read through LHAPDF~6~\cite{Buckley:2014ana}), the strong couplings of the shower and MPI, the MPI parameters and, for CP5-CR1, the string-fragmentation parameters. The tune settings are taken from the CMS software, with one compatibility setting for \textsc{Pythia}~8.318 (\cref{app:generator}). The test has two samples of each tune and one Monash control at a random top-specific strength $s\sim\mathcal{U}[0.02,0.14]$, which checks for false alarms and keeps the tunes blinded. Masses were drawn from $\{172.0,172.5,173.0\}\GeV$. Each sample has $7\times10^5$ events, generated in seven chunks with independent seeds and merged.

The frozen v3 package, the decision rule, the threshold and the fallback envelope were unchanged from oof12. The list of samples, the interpretation and the generator were fixed and hashed before any event was generated. Production could start only after a test showed that the new generator reproduces the old one event by event on a Monash sample. Predictions were hashed before the key was opened (\cref{app:blinding}).

\paragraph{Results} \Cref{tab:cms7} lists every sample.

\begin{table*}[tbp]
  \centering
  \caption{Per-sample results of the CMS-tune stress test. ``Mass-only shift'' is the $\mlb$ template fit (Monash templates) minus the true mass. The reported interval is the fallback for flagged samples and the primary interval otherwise. $\chimin$: goodness-of-fit statistic of the flow features (\cref{sec:inference}).}
  \label{tab:cms7}
  \small
  \begin{tabular}{llrrccr}
    \toprule
    Sample & True model & $\mt$ [GeV] & $\chimin$ & Flagged & Covers 95\% & Mass-only shift [GeV] \\
    \midrule
    cms\_04 & CP5 & 172.5 & 687 & yes & no & $+0.67$ \\
    cms\_05 & CP5 & 172.0 & 598 & yes & no & $+0.51$ \\
    cms\_01 & CP5-CR1 & 173.0 & 3924 & yes & no & $+0.54$ \\
    cms\_02 & CP5-CR1 & 172.0 & 3843 & yes & no & $+0.60$ \\
    cms\_03 & CP5-CR2 & 172.0 & 5635 & yes & yes & $+0.43$ \\
    cms\_07 & CP5-CR2 & 173.0 & 5594 & yes & no & $+0.66$ \\
    \midrule
    cms\_06 & Monash, $s=0.126$ & 172.5 & 4.8 & no & yes & $-0.53$ \\
    \bottomrule
  \end{tabular}
\end{table*}

\begin{itemize}
  \item \emph{Detection.} All six CMS samples exceed the threshold of 36.67 by a factor of 16 or more. CP5 gives the smallest $\chimin$ (600--690), CP5-CR1 about 3900 and CP5-CR2 about 5600. Had the narrow primary intervals been reported, they would have been off by $+0.56$ to $+1.65\GeV$.
  \item \emph{Coverage.} The fallback intervals covered 1 of 6 CMS samples at 95\%, and the true mass lay $0.02$--$0.18\GeV$ below their lower edges. The fallback assumes a CR-induced shift of at most $+0.30\GeV$, while the CMS tunes shift the fit by $+0.43$ to $+0.67\GeV$. The prespecified Gaussian comparator $G_{\mathrm{ens}}$ also covered only 1 of 7 samples. The evidence points to the Monash-based calibration and its uncertainty envelope as the source of the failure, the fallback is part of the reported procedure, and the procedure as a whole failed its intended coverage test.
  \item \emph{Control.} The Monash control was not flagged, its primary 95\% interval covered the truth, and the recovered strength was $\hat s=0.1225$ for a true $s=0.1263$.
\end{itemize}

The pre-registered verdict is therefore L: no critical failures (no unflagged CMS sample was miscovered), but five fallback misses. A warning is not a calibrated uncertainty, and this test shows the difference. The method recognised that the CMS tunes lie outside its training family, but the interval it then reported was too narrow for a nominal 95\% statement.

\paragraph{Generator-tune shift} The mass-only fit, calibrated on Monash, is shifted relative to the true mass by
\begin{equation}
  \begin{gathered}
  +0.59\pm0.07\GeV\ (\text{CP5}),\\ +0.57\pm0.07\GeV\ (\text{CP5-CR1}),\\
  +0.55\pm0.07\GeV\ (\text{CP5-CR2}),
  \end{gathered}
\end{equation}
where the uncertainties are statistical, from the single-sample resolution of $0.095\GeV$. The tune-level differences relative to CP5, $-0.02\pm0.10\GeV$ for CP5-CR1 and $-0.05\pm0.10\GeV$ for CP5-CR2, are consistent with zero. These are differences between complete tunes, not isolated CR effects: CP5-CR1 and CP5-CR2 were tuned with their own MPI parameters, and the six samples were generated at different masses. An exploratory follow-up at matched mass, which also splits CP5 into its components, is described in \cref{sec:diag8}.

The lesson for the method is that the $\mlb$ templates, and with them any in-situ CR correction, must be calibrated within the generator setup in use. Trained on Monash, the method detects a change of tune but cannot correct for it.

\subsubsection{Origin of the tune dependence (exploratory)}
\label{sec:cms7-origin}

Two exploratory studies, generated after cms7 had been unblinded and detailed in \cref{app:diag8}, identify the cause. First, the offset is a calibration effect. Five independent CP5 samples fitted with CP5 templates close to $-0.02\pm0.06\GeV$, while the same samples fitted with the frozen Monash templates are shifted by $+0.53\GeV$. Across single blocks of CP5 settings the shift follows the change of the $b$-jet response, about $0.9\GeV$ per 1\%, as \cref{eq:jes,eq:dm-estimate} predict. Second, the flow-to-mass relation is not universal. Within CP5, top-specific CR still lowers the fitted mass ($-0.50\pm0.11\GeV$ at $s=0.15$), with a slope about 30\% smaller than in the Monash family. The complete CMS CR variants change the charged particles in almost the same direction as the trained CR model (whitened cosines 0.84--0.87, equivalent to $s\approx0.13$--$0.15$), yet change the mass by only $+0.05\pm0.05\GeV$ relative to CP5. A mass correction based on charged particles is therefore valid only for models in which the link between charged particles and mass has been established, and its calibration depends on the tune.

\section{Validation at next-to-leading order}
\label{sec:nlo}

All results so far use leading-order $\ttbar$ showered by \textsc{Pythia}. A matched NLO calculation changes the hardest emission and the radiation pattern around the $b$-jets, which is exactly where the $\mlb$ fit and the in-cone flow features are sensitive. This section asks whether the conclusions of \cref{sec:cp5a9} survive in a full MC@NLO chain. It answers three questions. Can that chain be calibrated on its own? Are the mass shifts caused by top-specific CR and by the final-state shower coupling unchanged? And do the corresponding changes of the charged particles keep their direction and size? The analysis and its validity criteria were fixed, audited and hashed before the production was generated (\cref{app:blinding}), and were run unchanged.

\subsection{Production}
\label{sec:nlo-prod}

\paragraph{Chain} Hard events are generated with MG5\_aMC@\allowbreak NLO~3.5.5~\cite{Alwall:2014hca} for $pp\to\ttbar$ at NLO in QCD, at $\sqrts=13.6\TeV$, with NNPDF3.1 NNLO parton densities~\cite{NNPDF:2017mvq} through LHAPDF~6.5.5~\cite{Buckley:2014ana}. The top quarks are decayed with MadSpin~\cite{Artoisenet:2012st} to the dilepton final state with spin correlations. The events are showered with \textsc{Pythia}~8.318 and CP5~\cite{CMS:2019csb} using the MC@NLO matching prescription~\cite{Frixione:2002ik}: fourteen settings change relative to the LO CP5 samples, namely the Les Houches event (LHE) input in place of the internal hard processes and the MC@NLO shower settings (initial-state starting scale from the hard event, global recoil for the first final-state emission, matrix-element corrections off, \texttt{TimeShower:nPartonsInBorn = 2}), and the merge procedure verifies that nothing else changes. All samples use the \textsc{Pythia}~8.318 default recoil scheme for radiation in top decays, recoil-to-top (RTT), its effect is studied in \cref{sec:recoil}. The selection, the observables and the $\mlb$ fit are those of the LO study, with signed event weights in every histogram, mean and quantile.

\paragraph{Samples} Twenty-two MadGraph runs of $3.5\times10^5$ events each provide templates at $\mtmc=171.5$, 172.0, 173.0 and $173.5\GeV$ (three runs each) and at $172.5\GeV$ (six runs), and two independent closure samples at 172.5 and $172.25\GeV$ (two runs each) that never enter the templates. Top-specific CR (mode~3, $s=0.15$) and the Monash final-state radiation (FSR) coupling ($\alpha_s=0.1365$, one-loop running) shower the same six $172.5\GeV$ files as the nominal sample, so the three are paired event by event, the merge requires identical weights and identical hard-process spin observables on every shared event. In total $1.19\times10^7$ events were showered. The fraction of negative weights is 0.229 in every sample, so the effective sample size is $N_{\rm eff}\approx0.29\,N$: about $9.4\times10^4$ effective selected events at $172.5\GeV$ and in each paired variation, $4.7$--$4.8\times10^4$ at each other template mass and $3.2\times10^4$ in each closure sample. Four of the $2.1\times10^6$ events of the CR sample were rejected by \textsc{Pythia} after repeated beam-remnant failures, they are absent, and the event numbering of every remaining row was verified against the LHE file. MadSpin and the shower leave the spin correlation unchanged: the hard-process $\langle\cos\varphi\rangle$ and its value recomputed after the shower agree to within 0.001 in every sample.

\paragraph{Statistical treatment} Uncertainties come from a Poisson bootstrap with 300 replicas. In the paired samples one multiplicity is drawn per LHE event and shared by the nominal, CR and FSR showers, so selection migrations are included, the templates are rebuilt in every replica. The validity criteria, fixed in advance, require closure within two standard deviations with an acceptable fit (V1), no fit at the edge of the mass grid, no extrapolated response fit and template normalisation corrections below 1\% (V2), at most 1\% failed bootstrap replicas (V3), and a leave-one-file-out (LOFO) mean residual compatible with zero (V4). All four pass, with no failed replica.

\subsection{Calibration}
\label{sec:nlo-calib}

\Cref{fig:nlocalib} (left) compares the $\mlb$ distribution of the two chains at $172.5\GeV$. The NLO templates, interpolated linearly in $\mtmc$ bin by bin, describe the five template masses with $\chi^2=93.1$ for 78 degrees of freedom. The two independent closure samples give residuals of $-0.29\pm0.17$ and $-0.07\pm0.17\GeV$, or $-0.18\pm0.12\GeV$ combined, with fit qualities $\chi^2=24.5$ and 23.5 for 24 degrees of freedom (\cref{fig:nlocalib}, right). The calibration therefore closes within its statistical precision. This precision does not bound a calibration bias at the $0.1\GeV$ level, and the difference between the two closure fits, $+0.03\pm0.22\GeV$ for a true $+0.25\GeV$, shows that a quarter-GeV step is not resolved by samples of this size. The LOFO fits, in which each of the six $172.5\GeV$ files is fitted with templates built without it, have a mean residual of $+0.095\pm0.082\GeV$, one file lies at $+0.64\pm0.24\GeV$.

Fitted with the LO CP5 templates instead, the NLO samples are offset by $+1.01\GeV$ relative to the LO nominal, and three of the five masses fall outside the LO template range. This is a calibration-transfer offset, not an NLO shift of the top mass: like the tune dependence of \cref{sec:cp5a9}, it shows that the templates must be built in the chain in which they are used.

\begin{figure*}[tbp]
  \centering
  \includegraphics[width=\textwidth]{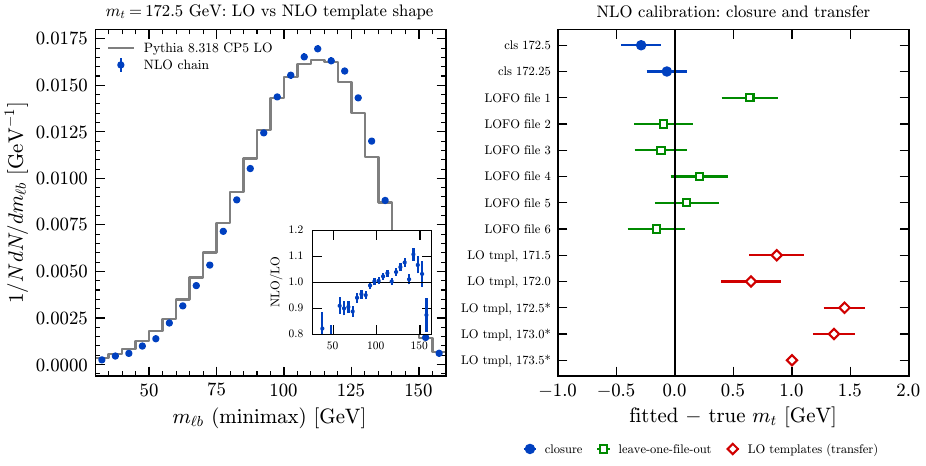}
  \caption{NLO calibration. Left: normalised minimax $\mlb$ at $\mtmc=172.5\GeV$ for the LO \textsc{Pythia} CP5 chain and the MG5\_aMC@NLO + MadSpin + MC@NLO-matched \textsc{Pythia} CP5 chain (signed weights), inset: NLO/LO, with the approximate statistical uncertainties of both chains combined in quadrature. Right: fitted minus true $\mtmc$ with NLO templates built from the five template masses. Filled circles: the two closure samples, independent of the templates ($-0.29\pm0.17$ and $-0.07\pm0.17\GeV$, combined $-0.18\pm0.12\GeV$). Open squares: leave-one-file-out fits of the six $172.5\GeV$ files (mean $+0.095\pm0.082\GeV$), an internal diagnostic. Open diamonds: NLO samples fitted with the LO templates (mean offset $+1.01\GeV$), a calibration-transfer offset rather than a physical shift, fits marked * lie outside the LO template range. The combined closure residual is compatible with zero, it does not bound the calibration bias to $0.12\GeV$.}
  \label{fig:nlocalib}
\end{figure*}

\subsection{Mass responses}
\label{sec:nlo-mass}

\begin{table*}[tbp]
  \centering
  \caption{Shift of the $\mlb$ template-fit mass at $\mtmc=172.5\GeV$, relative to the nominal sample of the same chain. LO: LO CP5 chain with LO templates. NLO: MC@NLO chain with NLO templates, paired samples. The templates are rebuilt in every bootstrap replica, so the uncertainties include template statistics.}
  \label{tab:nlomass}
  \footnotesize
  \setlength{\tabcolsep}{5pt}
  \begin{tabular}{lcccc}
    \toprule
    Variation & LO [GeV] & NLO [GeV] & NLO $-$ LO [GeV] & NLO $\chi^2$/d.o.f. \\
    \midrule
    Top-specific CR (mode 3, $s=0.15$) & $-0.555\pm0.114$ & $-0.480\pm0.108$ & $+0.075\pm0.156$ & 25.5/24 \\
    Monash FSR coupling & $-0.545\pm0.099$ & $-0.610\pm0.113$ & $-0.065\pm0.146$ & 22.5/24 \\
    \bottomrule
  \end{tabular}
\end{table*}

Both responses are significant in the NLO chain, at $4.4$ and $5.4$ standard deviations, and both agree with LO to within half a standard deviation (\cref{tab:nlomass,fig:nloresp}). The differences between LO and NLO, $+0.08\pm0.16$ and $-0.07\pm0.15\GeV$, are compatible with zero. About $0.15\GeV$ is the statistical precision of this comparison, it is not an upper limit on the difference. The LO values differ slightly from those of \cref{tab:cp5a9} ($-0.50$ and $-0.49\GeV$) because here the LO reference is recomputed with the same estimator as the NLO chain: two CP5 replicas as nominal and templates rebuilt in each bootstrap replica.

\begin{figure}[tbp]
  \centering
  \includegraphics[width=\columnwidth]{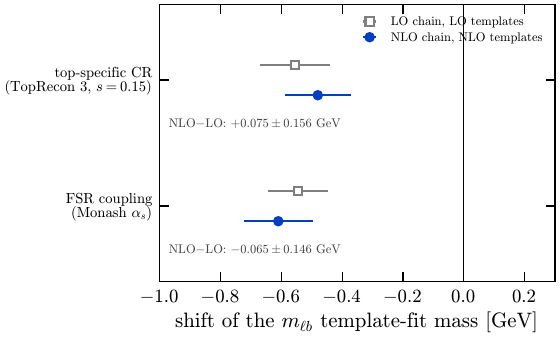}
  \caption{Shift of the $\mlb$ template-fit mass for top-specific CR (TopRecon mode 3, $s=0.15$) and for the Monash FSR coupling, each relative to the nominal sample of its chain. LO chain with LO templates: $-0.555\pm0.114$ and $-0.545\pm0.099\GeV$. NLO chain with NLO templates (paired samples, templates rebuilt in each bootstrap replica): $-0.480\pm0.108$ and $-0.610\pm0.113\GeV$. NLO $-$ LO: $+0.075\pm0.156$ and $-0.065\pm0.146\GeV$.}
  \label{fig:nloresp}
\end{figure}

\subsection{Charged-particle flow}
\label{sec:nlo-flow}

We compare the changes of the nine charged-particle observables as vectors, after whitening them with the covariance of the LO nominal sample, as in the frozen analysis. The direction of the CR response is the same in the two chains (cosine $0.998\pm0.003$), and so is its whitened size (NLO/LO $=0.98\pm0.03$). The same holds for the FSR response (cosine $0.989\pm0.006$, size ratio $1.03\pm0.04$). CR and FSR stay clearly distinct: their whitened cosine is $0.42\pm0.03$ at LO and $0.44\pm0.03$ at NLO, about $64^\circ$ apart. The six-feature version without the $\ptinb$ quantiles gives the same values to within 0.002. \Cref{fig:nloflow} shows the per-feature shifts.

The absolute flow of the nominal sample, by contrast, differs strongly between the chains: $\chi^2=152.9$ for nine features, driven by the in-cone features ($\langle\Ninb\rangle$ $-4.9\sigma$, $\langle\ptinb\rangle$ $-8.4\sigma$, $\ptinb$ 90\% quantile $-6.7\sigma$), while the UE features move by less than $2\sigma$. The \emph{changes} of the charged particles carry over from LO to NLO, but their absolute \emph{values} do not. Determining the CR strength from absolute values therefore needs charged-particle templates from the same chain as the mass calibration. This also applies to the comparison with data in \cref{sec:rivet}.

\begin{figure*}[tbp]
  \centering
  \includegraphics[width=\textwidth]{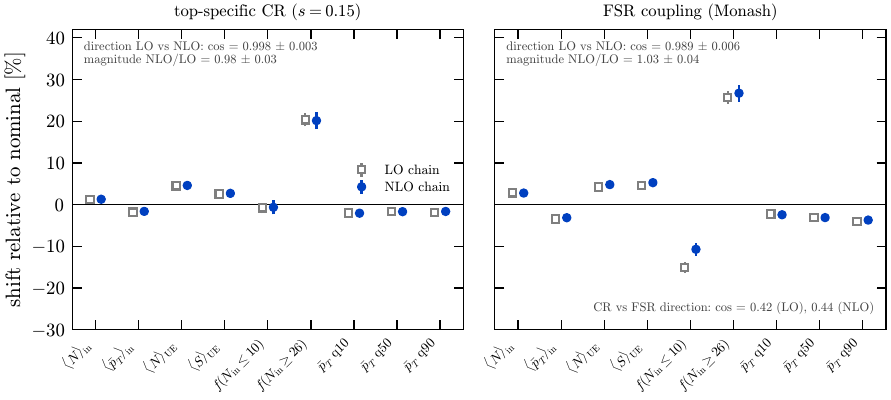}
  \caption{Response of the nine charged-flow features to top-specific CR (left) and to the FSR coupling (right), shown as percentage shifts relative to each chain's own nominal prediction (bootstrap uncertainties, paired for NLO). The quoted direction cosines and size ratios are computed from the full nine-feature response vectors whitened with the covariance of the LO nominal flow vector, not from the plotted percentages. The nominal flow differs strongly between the chains ($\chi^2=152.9$ for 9 features), while the CR and FSR response directions agree between LO and NLO (cosines $0.998\pm0.003$ and $0.989\pm0.006$). CR and FSR stay distinct (whitened cosine 0.42 at LO, 0.44 at NLO).}
  \label{fig:nloflow}
\end{figure*}

\subsection{Diagnostics and what the NLO check establishes}
\label{sec:nlo-diag}

Two features of the production were examined after the frozen result had been obtained, they do not change it. First, the LOFO outlier: the excess in $\langle\mlb\rangle$ of the first $172.5\GeV$ file appears only in the nominal shower and not in the CR and FSR showers of the same LHE events, which points to a shower-level fluctuation rather than a property of that MadGraph run, and across all independent LHE files the per-file $\langle\mlb\rangle$ are consistent ($\chi^2=16.5$ for 15 degrees of freedom). Second, the jet multiplicity varies between MadGraph runs more than expected ($\chi^2=30.8$ for 15 degrees of freedom), led by one run that is high in all three showers. The $\mlb$ means show no corresponding effect, and pairing the variations on common LHE events strongly reduces run-dependent effects in the responses, but does not guarantee that they cancel completely, because the variations change the jet activity and therefore the selection. We list this as a limitation to monitor (\cref{sec:outlook}).

In summary, the MC@NLO chain can be calibrated on its own: the combined closure residual on independent events is $-0.18\pm0.12\GeV$, about 1.5 standard deviations from zero, which does not establish closure at the 0.1\,GeV level. The mass responses to top-specific CR and to the final-state coupling differ between LO and NLO by amounts compatible with zero at the statistical precision of about $0.15\GeV$. Their flow signatures keep direction and size and remain distinguishable. Not established here are closure at the $0.1\GeV$ level, which needs about four times more closure statistics, the dependence on the recoil scheme (\cref{sec:recoil}), the dependence on the hadronisation model, and the agreement with data (\cref{sec:rivet}).

\section{Recoil scheme and \texorpdfstring{$b$}{b}-quark fragmentation}
\label{sec:recoil}

The flow correction of \cref{sec:inference} assumes that a change in the in-cone flow signals a change in CR. Two other parts of the $b$-jet modelling act on the same particles: the kinematic recoil of radiation off the $b$ quark in the top decay, and the fragmentation of the $b$ quark. Both are known to move top-mass observables. Public ATLAS studies have shown that the choice of recoil scheme affects radiation in top-quark decays, $b$-hadron fragmentation and $b$-jet observables, with consequences for reconstructed top-quark masses~\cite{ATLAS:2023mlb,ATLAS:PUB2025036,ATLASRecoil2026}. The recoil prescription must therefore be considered together with the final-state shower and fragmentation parameters. This section measures, in the NLO chain of \cref{sec:nlo}, how the two move the $\mlb$ fit and the flow, and what a CR-based flow correction would do to them.

\paragraph{Design} Two further samples shower the same six $172.5\GeV$ LHE files as the nominal, CR and FSR samples of \cref{sec:nlo}, so all five are paired event by event. Each changes exactly one setting on top of CP5 with MC@NLO matching. The first changes the treatment of radiation off the $b$ quark in the top decay. In the \textsc{Pythia}~8.318 default (\texttt{TimeShower:recoilStrategyRF = 0}), used in all other NLO samples, the $W$ boson takes the kinematic recoil, and separately the emission probability is reweighted (\texttt{TimeShower:weightRF = 1}) towards the eikonal radiation pattern of the top-associated dipole, the recoiler choice and the radiation pattern are thus set independently. We call this RTT, the customary shorthand for ``recoil to top''~\cite{Pythia8318doc}. The older option \texttt{recoilStrategyRF = 1}, recoil to the $b$ quark (RTB), was the default when CP5 was tuned. The \textsc{Pythia} authors have advised against using RTB as a recommended modern variation because of its effect on $b$ fragmentation~\cite{PythiaRecoil2026}, we use it as a diagnostic stress test of an older prescription, not as one endpoint of an uncertainty envelope. The second sets the Bowler parameter to $r_b=1.05$ (CP5: 0.855), the value of an ATLAS retuning of the Bowler parameter to LEP and SLD $b$-fragmentation data in the A14 configuration ($r_b=1.05\pm0.02$)~\cite{Muskinja:2024hf}. On CP5 it is a stress test of $b$-fragmentation, not a tune. The analysis uses the definitions of the frozen NLO analysis and was fixed and hashed before the samples were showered. As required, it reproduces the CR and FSR results of \cref{sec:nlo} exactly. Its validity criteria (intact pairing, at most 1\% failed bootstrap replicas, no fit at the grid edge or outside the template range) are all met. Both samples ran without generator failures, with $2.29\times10^5$ and $2.31\times10^5$ selected events shared with the nominal sample.

\begin{table*}[tbp]
  \centering
  \caption{Recoil-to-$b$ (RTB) and Bowler $r_b=1.05$ compared with top-specific CR and the FSR coupling, all in the NLO chain at $\mtmc=172.5\GeV$, paired with the nominal sample (RTT, $r_b=0.855$). Mass shifts use the NLO templates. Flow quantities use the nine features whitened, i.e.\ transformed with the inverse square root of the bootstrap covariance of the NLO nominal flow vector, ``cosine'' is the cosine of the angle between two such flow changes, and the $\chi^2$ outside the CR--FSR plane measures the part of a flow change that neither CR nor FSR explains. Residual mass error: the mass shift minus the correction implied by reading the flow change as CR only (projection $a_1$ on the CR direction), or as a combination of CR and FSR. The CR and FSR uncertainties differ slightly from \cref{tab:nlomass} because this analysis uses its own bootstrap.}
  \label{tab:recoil}
  \footnotesize
  \setlength{\tabcolsep}{3pt}
  \begin{tabular}{lcccc}
    \toprule
    & Top CR ($s=0.15$) & FSR coupling & RTB & $r_b=1.05$ \\
    \midrule
    Mass shift [GeV] & $-0.480\pm0.121$ & $-0.610\pm0.118$ & $+0.250\pm0.104$ & $-0.160\pm0.106$ \\
    \midrule
    Flow size rel.\ to CR & 1 & --- & $0.53\pm0.03$ & $0.50\pm0.02$ \\
    Cosine with CR & 1 & --- & $+0.25\pm0.05$ & $+0.19\pm0.05$ \\
    Cosine with FSR & --- & 1 & $+0.62\pm0.04$ & $+0.68\pm0.03$ \\
    $\chi^2$ outside CR--FSR plane & --- & --- & 487/7 & 413/7 \\
    \midrule
    CR-only projection $a_1$ & --- & --- & $0.13\pm0.03$ & $0.09\pm0.03$ \\
    Residual mass, CR-only corr. [GeV] & --- & --- & $+0.31\pm0.10$ & $-0.12\pm0.10$ \\
    Residual mass, CR+FSR corr. [GeV] & --- & --- & $+0.50\pm0.10$ & $+0.09\pm0.10$ \\
    \bottomrule
  \end{tabular}
\end{table*}

\paragraph{Mass} RTB raises the fitted mass by $+0.250\pm0.104\GeV$ relative to RTT, for the tested configuration, ATLAS studies of the recoil scheme in top-mass observables are reported in Refs.~\cite{ATLAS:2023mlb,ATLAS:PUB2025036}. Raising $r_b$ to 1.05 lowers it by $-0.160\pm0.106\GeV$, a $1.5\sigma$ effect (\cref{tab:recoil}, \cref{fig:recoil}, right). Both are comparable in size to the CR and FSR responses.

\paragraph{Flow} Both changes act inside the $b$-jets: they raise $\langle\Ninb\rangle$ (by $20$ and $19$ standard deviations of the paired comparison) and lower $\langle\ptinb\rangle$ ($-14\sigma$ each), so the $b$-jets contain more, softer charged particles (\cref{fig:recoil}, left). The UE outside the jets is almost unchanged, by at most $3.5\sigma$ or about 1\%, whereas top-specific CR and the FSR coupling raise it by 3--5\%. Measured in the whitened nine-observable space, both changes are about half the size of the CR change and nearly perpendicular to it (cosines 0.25 and 0.19), and closer to the FSR direction (cosines 0.62 and 0.68). Most of each change cannot be described as a combination of CR and FSR (remaining $\chi^2$ of 487 and 413 for 7 degrees of freedom). This $\chi^2$ is only approximate, because its covariance comes from the bootstrap with the decomposition refitted in every replica. These values come from the analysis version fixed in advance for the official result, an earlier version with a different covariance was replaced before these samples existed and was not used (an independent reproduction with it gives 432/7 and 369/7).

\paragraph{Consequence for a flow correction} Read as CR only, the RTB flow change implies a small correction ($a_1=0.13$), and the mass error left after it is $+0.31\pm0.10\GeV$: the flow correction neither removes nor creates a large bias. Reading the flow change as a combination of CR and FSR makes it worse, $+0.50\pm0.10\GeV$, because the FSR-like part of the RTB change is attributed to an FSR change with the opposite mass sign. For $r_b=1.05$ the two readings leave $-0.12\pm0.10$ and $+0.09\pm0.10\GeV$. These numbers describe a hypothetical linear-response correction, not an end-to-end inference. They show that adding FSR as a nuisance direction without a recoil or fragmentation direction is not safe, and the large residuals outside the CR--FSR plane suggest that such a mismatch would be detectable, end-to-end detection by the goodness-of-fit test has not been validated for these variations.

\paragraph{Near alignment (post hoc)} After the pre-registered result, we compared the two flow changes directly. They point in almost the same direction (whitened cosine $0.96\pm0.01$ in an independent bootstrap, a decomposition of RTB on CR, FSR and $r_b$ gives coefficients $0.03$, $-0.05$ and $1.10$ with a residual of 28\% of its whitened length) and have similar size (0.53 and 0.50 of the CR response), while their central mass shifts have opposite signs (the $r_b$ shift is only $1.5\sigma$ from zero). A cosine of 0.96 establishes strong alignment, not statistical indistinguishability: the whitened difference of the two vectors is about 0.15 of the CR response, and whether it is resolvable requires a direct separation test with the covariance of the difference, which we have not performed. In this feature set the two effects are therefore potentially difficult to disentangle, and the flow cannot be relied on to calibrate the mass for them, that needs an external constraint, such as $b$-fragmentation measurements in $e^+e^-$ or $\ttbar$ events, or additional features (\cref{sec:outlook}). With this independent bootstrap the RTB--CR cosine is $0.14\pm0.04$ rather than $0.25\pm0.05$: differences of about 0.1 between small whitened cosines depend on the bootstrap estimate of the $9\times9$ covariance and are not significant.

\begin{figure*}[tbp]
  \centering
  \includegraphics[width=\textwidth]{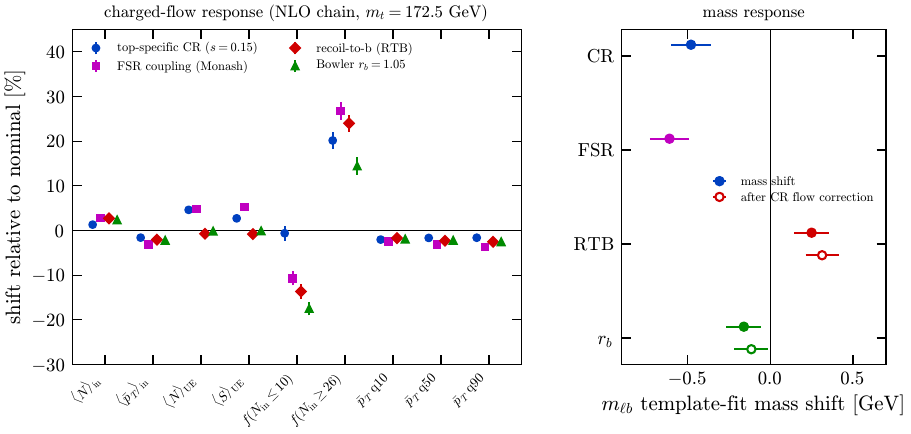}
  \caption{Recoil scheme and $b$-fragmentation in the NLO chain at $\mtmc=172.5\GeV$, all samples paired with the nominal one (RTT, $r_b=0.855$). Left: percentage shift of the nine flow features for top-specific CR ($s=0.15$), the Monash FSR coupling, recoil-to-$b$ (RTB) and Bowler $r_b=1.05$ (paired bootstrap uncertainties, illustration, the whitened quantities of \cref{tab:recoil} are computed from the full vectors). Right: shift of the $\mlb$ template-fit mass (filled) and, for RTB and $r_b$, the residual mass error after a CR-only flow correction (open).}
  \label{fig:recoil}
\end{figure*}

\paragraph{Implications} In the tested configurations, recoil and $b$-fragmentation each shift the $\mlb$ fit by $0.2$--$0.3\GeV$, and a CR correction does not capture them. For an experimental analysis this means three things: state the recoil scheme used in the calibration (here RTT) and treat recoil as a separate modelling effect, using the variations recommended for current shower versions, constrain $b$-fragmentation with external measurements, and use the goodness-of-fit test to detect, not to correct, changes that do not look like CR. A useful property for the comparison with data follows from the UE result: since RTB and $r_b$ leave the particles outside the jets almost unchanged while CR moves them, a constraint on the CR strength from the ATLAS measurement of exactly those particles~\cite{ATLAS:2022zbu} should be largely insensitive to the recoil scheme and $r_b$, \cref{sec:rivet} tests this with the RTB sample.

\section{Comparison with the ATLAS measurement}
\label{sec:rivet}

ATLAS has measured the multiplicity $n_{\rm ch}$ and the scalar transverse-momentum sum $\sum\pT$ of charged particles outside jets in $\emu$ events with two $b$-jets at $\sqrts=13\TeV$, unfolded to particle level, with $139\ifb$~\cite{ATLAS:2022zbu}. These particles are the underlying-event part of our flow features. \Cref{sec:recoil} shows that they respond to top-specific CR (by 3--5\%) but hardly to the recoil scheme or to $r_b$ (about 1\%). The measurement is therefore a natural place to ask which CR strengths the data allow within CP5, and what mass shift that range corresponds to. The answer is conditional on CP5 and on the TopRecon model, it is not a tune.

\subsection{Set-up}
\label{sec:rivet-setup}

\paragraph{Samples} Three MG5\_aMC@NLO runs of $3.5\times10^5$ events at $\sqrts=13\TeV$ and $\mtmc=172.5\GeV$, with the cards of \cref{sec:nlo} apart from the beam energy, are showered five times with \textsc{Pythia}~8.318 CP5 and MC@NLO matching: without top-specific CR ($s=0$), with TopRecon mode 3 at $s=0.15$, 0.30 and 1.0, and with recoil-to-$b$ at $s=0$. All five are paired event by event.

\paragraph{Rivet} The official routine \texttt{ATLAS\_2022\_I2152933} of Rivet~3.1.8~\cite{Bierlich:2019rhm} runs in the same process as the shower, on every generated event, with Rivet's beam-energy check enabled. A dedicated validation verified that the event record passed to Rivet leaves the shower output unchanged, that Rivet receives every event with its signed weight (its own event counter agrees with \textsc{Pythia}'s sums of $w$ and $w^2$), and that the YODA output is read correctly. In the corrected 13\,TeV CP5 sample the Rivet selection accepts 11.9\% of the events, in the leading-order validation sample it accepted 12.6\%, against 12.9\% for the selection of \cref{sec:setup}.

\paragraph{Predictions and $\chi^2$} Predictions are built from Rivet's unnormalised histograms summed over the three files and normalised exactly like the data, the one-dimensional normalisation must reproduce Rivet's own to $10^{-6}$. The MC statistical covariance is propagated from the signed weights, including the normalisation,
\begin{equation}
  \mathrm{Cov}(f_i,f_j)=\frac{\delta_{ij}S_i-f_iS_j-f_jS_i+f_if_j\sum_kS_k}{W^2},
  \label{eq:mccov}
\end{equation}
with $f_i=W_i/W$ the bin fractions and $S_i$ the sums of squared weights. The $\chi^2$ uses the data covariance plus \cref{eq:mccov}, with the last bin of every normalisation group removed.

\paragraph{Pre-registered rules} Before any 13\,TeV prediction existed we fixed: the primary observable ($n_{\rm ch}$), a sensitivity rule (expected $\Delta\chi^2$ between $s=0$ and 0.15 of at least 4), the $s$ fit (piecewise-linear interpolation on $s\in[0,1]$, intervals where $\chi^2\le\chi^2_{\min}+1$ and $+4$, if the first reaches a boundary, a one-sided bound at $\chi^2_{\min}+2.71$ is quoted instead). Under the usual asymptotic assumptions these correspond to about 68.3\%, 95.4\% and 95\% one-sided, exact frequentist coverage near the physical boundary $s=0$ has not been validated, and the thresholds are reported as fixed in advance rather than adjusted, a goodness-of-fit rule (for $p<0.01$ the result is only a \emph{preference within CP5}), a recoil check with the RTB sample, an energy check of the mass response against \cref{sec:nlo}, and the translation of the $s$ range into a mass shift.

\subsection{Validation of the published inputs}
\label{sec:rivet-hepdata}

Every covariance matrix of the HEPData record~\cite{ATLAS:2022zbu} was required to be symmetric, positive definite after removing the last bin of each normalisation group, and to satisfy the normalisation constraint $\sum_jC_{ij}\Delta_j=0$. Several matrices fail (\cref{tab:hepdata} in \cref{app:hepdata}). In the $n_{\rm ch}$ covariance that includes theory uncertainties, three pairs of entries that should be equal ($C_{ij}$ and $C_{ji}$) differ, by factors of 10 and 1000 or because one entry is copied from a neighbouring one, and the matrix has a negative eigenvalue. The exact entries and eigenvalues are listed in \cref{app:hepdata}, and the checks are reproducible with the archived scripts. Two blocks of the double-differential covariances are labelled with the bins of another block, since the values of their mirror blocks are intact, they are replaced by the transposes. These are inconsistencies in the public entries of record version~1 as downloaded for this study, they have not been confirmed by the HEPData maintainers or the analysis authors, to whom they will be reported. The main comparison therefore uses the $n_{\rm ch}$ covariance without theory uncertainties, which passes all checks, the published $n_{\rm ch}$ uncertainties also agree better with it. The goodness-of-fit $p$-values below depend on this choice of covariance.

A second convention issue concerns the routine itself. The double-differential data integrate to unity with an $n_{\rm ch}$ slice width of 20 for every slice, while the routine divides each slice by its own edges (19.5, 20, 20, 20 and 21.5). Rivet's finalised first and last slices are therefore off by $20/19.5$ and $20/21.5$ relative to the data, we build the predictions with the data convention.

\subsection{Particle-level stability convention}
\label{sec:rivet-tau}

The measurement defines its particle level with ``stable particles with a mean lifetime $\tau>30$\,ps''~\cite{ATLAS:2022zbu}, so that $K^0_S$, $\Lambda$, $\Sigma^\pm$, $\Xi$ and $\Omega$ hadrons are stable and their decay products are not counted. Our first pre-registered 13\,TeV production used the \textsc{Pythia} default, in which these hadrons decay, this was found only after that result had been examined. In the NLO chain 8.6\% of the charged particles with $\pT>0.5\GeV$ and $|\eta|<2.5$, and 7.7\% of those outside jets, descend from such decays, and in the Rivet selection the paired productions show that they raise $\langle n_{\rm ch}\rangle$ by 5.1\% and $\langle\sum\pT\rangle$ by 3.7\%. That comparison is therefore not used. A corrected production differs only by \texttt{ParticleDecays:limitTau0 = on} (\texttt{tau0Max} = 10\,mm, $c\tau=10$\,mm corresponds to 33.4\,ps rather than exactly 30\,ps, and no relevant species lies between the two). Its analysis was fixed before the new events existed and is labelled as a correction made after the first result: the declared changes are the new inputs, the exclusion of $s=1.0$ from the mass translation (its mass fit lies outside the template range) and a measurement of the effect of the decay convention on the mass responses. The flow features of \crefrange{sec:fingerprint}{sec:recoil} use the \textsc{Pythia} default consistently in all samples, their definition is stated in \cref{sec:setup}.

\subsection{Mass responses at 13\,TeV}
\label{sec:rivet-mass}

The paired samples also give the top-mass responses at 13\,TeV (\cref{tab:mass13}), with the definitions and the NLO templates of \cref{sec:nlo}. The CR response at $s=0.15$ agrees with the 13.6\,TeV value (difference $+0.04\pm0.18\GeV$, the pre-registered energy check), and the RTB response with that of \cref{sec:recoil}. The response grows less than linearly with $s$, at $s=1.0$ the fit lies outside the template range and is not used. The stable-particle convention changes the responses by $+0.03$ to $+0.05\GeV$, with uncertainties of $0.20\GeV$, and the absolute fit by $-0.03\pm0.15\GeV$: the mass results of \crefrange{sec:nlo}{sec:recoil}, obtained with the \textsc{Pythia} default decays, are unaffected within their precision.

\begin{table*}[tbp]
  \centering
  \caption{Mass responses at $\sqrts=13\TeV$ relative to CP5 without top-specific CR, NLO chain, paired bootstrap. Corrected production with the ATLAS stable-particle convention, the first production, with the \textsc{Pythia} default decays, is shown for comparison.}
  \label{tab:mass13}
  \small
  \begin{tabular}{lccl}
    \toprule
    Variation & $\Delta m$ [GeV] & Default decays & Remark \\
    \midrule
    TopRecon mode 3, $s=0.15$ & $-0.44\pm0.14$ & $-0.47$ & 13.6\,TeV: $-0.48\pm0.11$ \\
    TopRecon mode 3, $s=0.30$ & $-0.64\pm0.15$ & $-0.69$ & \\
    TopRecon mode 3, $s=1.0$ & $-1.14\pm0.16$ & $-1.20$ & outside template range, not used \\
    Recoil-to-$b$ & $+0.42\pm0.14$ & $+0.39$ & 13.6\,TeV: $+0.25\pm0.10$ \\
    \bottomrule
  \end{tabular}
\end{table*}

\subsection{Result}
\label{sec:rivet-result}

The corrected production completed without failures beyond the three events per TopRecon sample also seen in the first production, and all merge checks passed: settings, event pairing within the production and with the first production, and Rivet's bookkeeping. The frozen analysis was then run unchanged. Its integrity checks, mass fits and bootstrap all passed. The results below follow the pre-registered order, \cref{tab:rivetchi2} collects the $\chi^2$ values and \cref{fig:rivetdata,fig:rivetfit} show the distributions and the fit.

\paragraph{Sensitivity} The expected $\Delta\chi^2$ between $s=0$ and $s=0.15$, computed without the data values, is 24.3 for the primary $n_{\rm ch}$ comparison (28.6 with the default decays). It exceeds the threshold of 4, so the measurement is sensitive to top-specific CR at the reference strength. Every cross-check configuration is sensitive as well (7.3 to 61 at $s=0.15$), while the RTB sample differs from $s=0$ by only 1.1 to 6.7.

\begin{table*}[tbp]
  \centering
  \caption{$\chi^2$ of the CP5 predictions against the ATLAS data at the generated points, with data and MC statistical covariances, the last bin of every normalisation group is removed. ``No theory'': covariance without the theory uncertainties of the measurement (table 57 for $n_{\rm ch}$, the primary configuration). RTB: recoil-to-$b$ shower option at $s=0$. Inconsistent covariances (\cref{tab:hepdata}) are not used. The global $n_{\rm ch}$ and $\sum\pT$ row is an approximate diagnostic: the MC statistical cross-covariance between the two separately normalised observables is not available from the histograms and is set to zero (the data cross-covariance is included).}
  \label{tab:rivetchi2}
  \footnotesize
  \begin{tabular}{lccccc c}
    \toprule
    Observable (covariance) & $s=0$ & $s=0.15$ & $s=0.30$ & $s=1.0$ & RTB & ndf \\
    \midrule
    $n_{\rm ch}$ (no theory) & 95.1 & 110.6 & 151.0 & 230.8 & 107.2 & 7 \\
    $\sum\pT$ (with theory) & 137.3 & 90.8 & 84.4 & 43.2 & 132.3 & 10 \\
    $\sum\pT$ (no theory) & 254.4 & 155.1 & 129.6 & 56.9 & 247.6 & 10 \\
    $n_{\rm ch}$ and $\sum\pT$, global (no theory) & 540.5 & 569.6 & 639.3 & 751.0 & 538.5 & 17 \\
    $\sum\pT$ in $n_{\rm ch}$ slices (no theory) & 562.5 & 856.9 & 1083.7 & 1490.7 & 554.9 & 8 \\
    \bottomrule
  \end{tabular}
\end{table*}

\paragraph{Goodness of fit and the preferred strength} The fit of $s$ to the primary $n_{\rm ch}$ distribution gives $\chi^2_{\min}=93.4$ for 6 degrees of freedom ($p=6\times10^{-18}$). Under the pre-registered rule the result is therefore not a constraint on $s$ but a \emph{preference within CP5 and TopRecon}: $\hat s=0.036$, with $[0.008,0.064]$ at $\chi^2_{\min}+1$ and $[0,0.091]$ at $\chi^2_{\min}+4$. Because the model does not describe the data, these intervals understate the uncertainty. Scaling $\Delta\chi^2$ by $\chi^2_{\min}/\mathrm{ndf}=15.6$, an indicative and not pre-registered correction, widens the $1\sigma$ range to $[0,0.144]$, neither range has validated frequentist coverage. Cubic instead of linear interpolation between the generated points gives $\hat s=0.025$. The pre-registered statistic profiles $\chi^2(s)$ although the MC covariance depends on $s$, a Gaussian likelihood would add $\ln\det C(s)$. As a post-hoc robustness check we included this term: it varies by 0.05 over the scanned range, because the MC variances are small compared with the data variances (median ratio 6\%), and $\hat s$ stays at 0.036 with the interval endpoints changing by at most $10^{-4}$. This check does not modify the pre-registered statistic. For comparison, the first production with the \textsc{Pythia} default decays (\cref{sec:rivet-tau}) gave $\chi^2_{\min}=136.8$ for 6 degrees of freedom with the minimum at the boundary $s=0$, the decay convention improves the description but does not make it acceptable, and that first result is not used.

\paragraph{What the data show} CP5 already predicts more charged particles outside jets than measured: $\langle n_{\rm ch}\rangle=42.1$ against about 39.0 in the data, estimated from the bin centres. The measured distribution is also narrower, with more events at $20\le n_{\rm ch}<40$ and fewer above 60, where the prediction exceeds the data by up to a factor of three (\cref{fig:rivetdata}, left). Top-specific CR adds particles, so $n_{\rm ch}$ prefers small $s$: the observed $\chi^2$ rises from $s=0$ to $s=0.15$ by 15.6. The $\sum\pT$ distribution pulls the other way (\cref{fig:rivetdata}, right): CP5 predicts too many events at small $\sum\pT$ and too few around 40\,GeV, and its fit, a cross-check, runs to $s=1$ with $\chi^2_{\min}=57$ for 9 degrees of freedom. The two-dimensional cross-checks give $\hat s=0.038$ (global, approximate) and $\hat s=0$ (slices), with $\chi^2_{\min}$ of 537 and 563. No value of $s$ describes both distributions. The mismatch lies in the CP5 description of the underlying event in these events, which top-specific CR cannot repair, rather than in the CR strength itself.

\paragraph{Stability (post hoc)} Fits to the three LHE files separately give $\hat s=0.086$, 0.074 and 0.000. They differ by more than the pre-registered $1\sigma$ intervals, and each file alone also fails the goodness-of-fit criterion ($\chi^2_{\min}$ between 58 and 75 for 6 degrees of freedom). This is consistent with a poorly fitting profile being sensitive to sample fluctuations, a dedicated test of the distribution of the minimum has not been performed. All three lie inside the scaled interval, whose coverage is not known either.

\paragraph{Recoil check} Shifting every prediction by the RTB change moves $\hat s$ by $+0.021$. The RTB change is 0.35 of the $s=0.15$ change in the metric of the data, and RTB does not improve the fit ($\chi^2=107.2$ against 95.1). As expected from \cref{sec:recoil}, the particles outside jets constrain CR largely independently of the recoil scheme.

\paragraph{Mass translation} Within the calibrated range $s\le0.30$, the preferred strength corresponds to $\Delta m(\hat s)=-0.11\GeV$ relative to CP5 without top-specific CR (\cref{fig:rivetfit}, right). The $1\sigma$ range of $s$ maps onto $[-0.19,-0.02]\GeV$, and adding the MC statistical, interpolation and recoil terms in quadrature gives $[-0.26,+0.05]\GeV$. With the scaled interval the range is $[-0.42,0]\GeV$. These numbers describe what the data prefer within CP5 and one CR model, because the model fails the goodness-of-fit test, they are neither a validated data-driven mass correction nor a measurement of the CR contribution to the top-mass uncertainty.

\begin{figure*}[tbp]
  \centering
  \includegraphics[width=\textwidth]{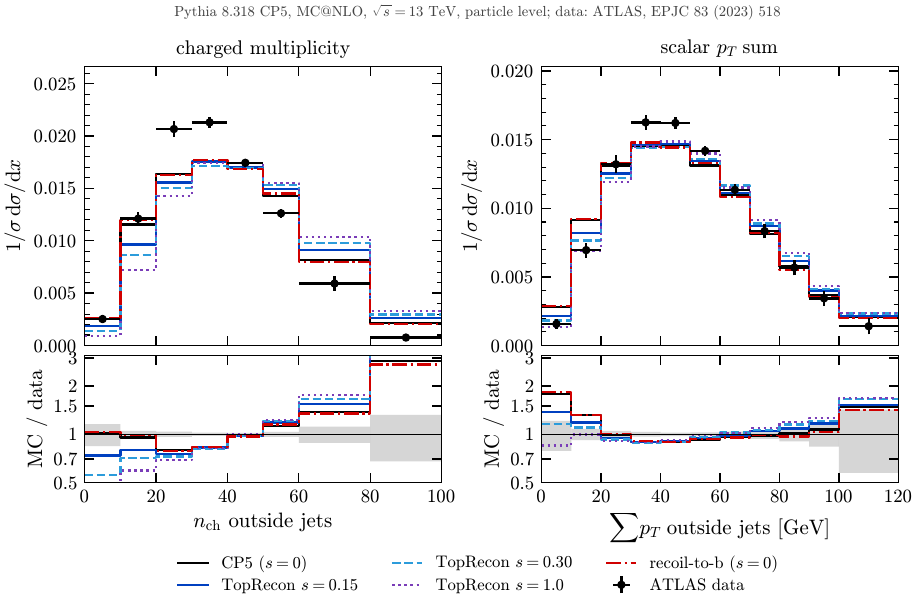}
  \caption{Normalised distributions of the charged-particle multiplicity (left) and scalar transverse-momentum sum (right) outside jets in $\emu$ events with two $b$-jets at $\sqrts=13\TeV$: ATLAS data~\cite{ATLAS:2022zbu} with the errors of the covariance without theory uncertainties, and \textsc{Pythia}~8.318 CP5 predictions in the MC@NLO chain with the ATLAS stable-particle convention, without top-specific CR, with TopRecon mode 3 at three strengths, and with recoil-to-$b$. Lower panels: prediction over data (log scale), the grey band is the data uncertainty.}
  \label{fig:rivetdata}
\end{figure*}

\begin{figure*}[tbp]
  \centering
  \includegraphics[width=\textwidth]{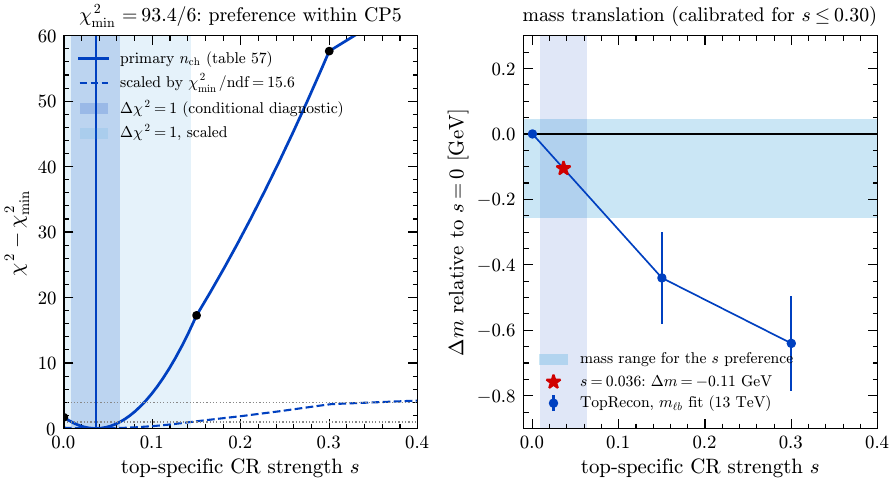}
  \caption{Left: $\chi^2-\chi^2_{\min}$ of the primary $n_{\rm ch}$ comparison as a function of the top-specific CR strength (piecewise-linear interpolation between the generated points, dots), unscaled (solid) and scaled by $\chi^2_{\min}/\mathrm{ndf}$ (dashed), with the corresponding $1\sigma$ ranges. These are conditional diagnostic $\Delta\chi^2$ contours, not confidence intervals with validated coverage, because CP5 fails the goodness-of-fit test. Right: mass response at 13\,TeV relative to $s=0$ with its uncertainties, the preferred strength (star), its $1\sigma$ range of $s$ (vertical band) and the resulting mass range including the MC, interpolation and recoil terms (horizontal band). The mass band is a diagnostic translation within CP5, not an uncertainty on the CR effect in a measured top-quark mass.}
  \label{fig:rivetfit}
\end{figure*}

\paragraph{Summary} The unfolded ATLAS observables have statistical sensitivity to top-specific CR at the reference strength, but a data-driven in-situ mass calibration has not been demonstrated with a model that describes the measured distributions. The CP5 underlying event in $\ttbar$ events is mismodelled, in opposite directions for $n_{\rm ch}$ and $\sum\pT$, and within CP5 the data only prefer a small CR strength, $\hat s\approx0.04$, corresponding to a mass shift of about $-0.1\GeV$. A constraint on the CR strength from these data needs an underlying-event model that describes them first, the limiting factor is the tune, not the sensitivity.

An independent replay of these calculations from the YODA files and the HEPData tables, including the paired-bootstrap mass responses, reproduces the numbers above. The HEPData extract used here is reproduced byte for byte from the record's YAML files by the archived extraction script, and the Rivet reference file matches its pinned hash.

\section{Limitations and outlook}
\label{sec:outlook}

The method works within the family it was calibrated on and detected every strong mismatch we tested, but its uncertainty model does not cover a change of generator tune, and several QCD effects on the $b$-jets are potentially difficult to disentangle with the current observables.

\paragraph{Limitations}
\begin{itemize}
  \item \emph{Level and generator.} All results are at particle level with \textsc{Pythia}~8 and a single string-fragmentation model. Detector effects smear both $\mlb$ and the track observables, the tracking nuisance is a count-level proxy that leaves $\ptinb$ unchanged. A cluster-hadronisation generator (\textsc{Herwig}~7) is untested.
  \item \emph{One CR model family.} The in-situ correction profiles top-specific CR (TopRecon). Complete CMS tune variants move the flow in a similar direction without moving the mass (\cref{sec:diag8}), so the flow-to-mass relation is not universal, and its calibration is tune dependent (\cref{sec:cp5a9}).
  \item \emph{Blind tests at LO.} The three blinded tests were run at leading order with Monash templates. The NLO study validates the responses and the calibration (\cref{sec:nlo}), no NLO-trained estimator has been tested blind for coverage, and all results are at generator level. The closure precision is $\pm0.12\GeV$, which cannot resolve a calibration bias at the $0.1\GeV$ level.
  \item \emph{Recoil and fragmentation.} The \textsc{Pythia}~8.318 default recoil treatment (RTT) is used throughout, CP5 was tuned with the older RTB option, which serves here as a stress test, not a validated modelling uncertainty. Recoil and $b$-fragmentation produce closely aligned flow changes with opposite central mass shifts (\cref{sec:recoil}), so they need external constraints. The flow-to-mass relation is not universal and no universal map is claimed.
  \item \emph{Fallback and coverage.} The $+0.30\GeV$ upper edge of the fallback envelope was exceeded by CR off ($+0.40$ to $+0.42\GeV$) and by all CMS tunes ($+0.43$ to $+0.67\GeV$). val24, oof12 and cms7 total 43 samples, all conditional on one training set, and the gain over a mass-only method has not been measured at matched coverage.
  \item \emph{Production heterogeneity.} The jet multiplicity varies between MG5\_aMC@NLO runs more than expected, pairing the variations on common events reduces, but does not guarantee the cancellation of, run-dependent effects.
  \item \emph{Comparison with data.} The ATLAS comparison is conditional on CP5 and on one CR model, and is made at $13\TeV$ while the rest of the study uses $13.6\TeV$. CP5 does not describe the data ($p=6\times10^{-18}$), so the result is a preference within CP5, not a constraint, and the single-file fits show that its minimum is not stable at the level of the pre-registered intervals. Its published covariance including theory uncertainties is not usable for $n_{\rm ch}$ (\cref{sec:rivet-hepdata}).
  \item \emph{Other hadronisation mechanisms.} Rope hadronisation and junction formation change the yields of strange hadrons and baryons with event activity~\cite{Bierlich:2015rha,Bierlich:2026ropes}. With the ATLAS stable-particle convention these hadrons enter the charged-particle counts, and these mechanisms are outside the trained family, only junction-forming CR was tested, through CP5-CR1 (\cref{sec:cms7}).
\end{itemize}

\paragraph{Outlook}
\begin{enumerate}
  \item An NLO-calibrated version of the estimator with separate nuisance directions for CR, the final-state shower, recoil and $b$-fragmentation, tested blind on independent samples against a mass-only method at matched coverage, with an output that states whether its uncertainty is supported.
  \item A joint fit of underlying-event parameters and the CR strength to the ATLAS data, so that the CR constraint is not conditional on one tune. \Cref{sec:rivet-result} shows that this is necessary: CP5 mismodels the multiplicity and the $\pT$ sum outside jets in opposite directions. Event reweighting of underlying-event and CR parameters~\cite{Pabst:2026rew} would make such a scan affordable.
  \item \textsc{Herwig}~7 with its cluster model and its CR variants, and external $b$-fragmentation constraints from $e^+e^-$ and $\ttbar$ measurements.
  \item A parametrised detector response for leptons, jets and tracks, including the robustness of track observables outside jets against pile-up. At the high-luminosity LHC (HL-LHC) the gain comes mainly from reducing systematic uncertainties and from vertexing, not from luminosity.
\end{enumerate}

\paragraph{Experimental use} All inputs are tracks and standard dilepton objects. In an experimental implementation, charged-particle flow could constrain the strength of a calibrated CR model and potentially reduce the corresponding component of the model-spread uncertainty. The remaining dependence on the CR model family would still require independent evaluation. Whether the gain survives at detector level, with real $b$-tagging and pile-up, and across generators, is the decisive open question.

\section{Conclusions}
\label{sec:conclusions}

Charged particles in and around the $b$-jets of dilepton $\ttbar$ events carry enough information to constrain colour reconnection in situ within a calibrated model family. In blinded leading-order tests, a frozen mass-plus-flow estimator reduced the root-mean-square mass error from $0.18\GeV$ for a mass-only fit to $0.096\GeV$, with coverage close to nominal. Out of family it flagged CR off and complete CMS tunes, but its fallback did not cover the CMS tunes, whose $\mlb$ calibration differs by $+0.4$ to $+0.7\GeV$: the method detects a mismatch it cannot quantify, and must be calibrated within the generator set-up in use.

The central responses are confirmed at next-to-leading order within the statistical precision. In a matched MC@NLO chain the calibration closure residual on independent samples is $-0.18\pm0.12\GeV$. The mass responses to top-specific CR and to the final-state shower coupling agree with leading order within about $0.15\GeV$, and their flow directions are unchanged. The absolute flow values, however, depend on the chain, so templates and data must be compared within one chain.

Two further mechanisms set the limits of the approach. Stress tests of the top-decay recoil scheme and of the $b$-quark fragmentation function move the fitted mass by $+0.250\pm0.104\GeV$ and $-0.160\pm0.106\GeV$. Both act on the same in-cone particles and have closely aligned flow signatures, although their central mass shifts have opposite signs, whether they can be separated statistically has not been tested. A CR-only flow correction neither removes nor creates a large bias for them, but it cannot calibrate them either, they require external constraints. The particles outside jets, in contrast, respond to CR but hardly to recoil or fragmentation, which makes the ATLAS measurement of exactly these particles a natural constraint on the CR strength within a given tune.

The comparison with that measurement confirms the sensitivity and exposes the limiting factor. The expected sensitivity of the predictions separates CR strengths that differ by 0.15, and the recoil scheme hardly affects the comparison. However, CP5 overpredicts the charged multiplicity outside jets and mismodels the shape of the $\pT$ sum, so that no CR strength describes both. Within CP5 the data prefer a small strength, $\hat s=0.036$, corresponding to a mass shift of about $-0.1\GeV$, this is a preference within one tune, not a constraint. Turning it into a constraint requires an underlying-event description that fits these data, for example a joint fit of underlying-event parameters and the CR strength, or a different generator.

The principal result is a prospective, model-conditional strategy for constraining a specified colour-reconnection parameter through charged-particle flow, together with a frozen blind-validation protocol that explicitly identifies its failures. The present implementation improves mass recovery within the trained model family, but it does not provide a calibrated uncertainty across independent generator tunes or a data-driven CR correction from the ATLAS measurement. The protocol also exposed a defect in our own first comparison with data, corrected openly, and inconsistencies in published covariance matrices that we found while validating the inputs.

\section*{Acknowledgments}
The work of A.~Bellagroudi is funded by the National Center for Scientific and Technical Research (CNRST) under the PhD-Associate Scholarship (PASS).

\section*{Data availability}
The analysis code, the simulation configurations (generator cards and random seeds), the frozen analysis packages with their hash manifests, the analysis log and the reduced results behind every table and figure are available from the corresponding author upon reasonable request. The experimental data used in this study are publicly available from the ATLAS HEPData record ins2152933 and through the Rivet routine \texttt{ATLAS\_2022\_I2152933}.

\section*{Declaration of competing interest}
The authors declare that they have no known competing financial interests or personal relationships that could have appeared to influence the work reported in this paper.

\appendix
\section{Blinding and reproducibility}
\label{app:blinding}

Each blind test followed the same four steps: freeze the method and hash its files, generate and anonymise the samples, with the key kept private, run the prediction script and record the SHA-256 of its output, only then open the key and run the scoring script. The tests enforced the last step differently. The oof12 and cms7 scoring scripts take the committed hash as an argument and refuse to run on a mismatch. The val24 scoring script has no such check: for val24 the prediction hash was recorded in the unblinding log before the key was received, and the match was verified by hand before scoring. The SHA-256 hashes of the frozen inference package, the pre-registrations and the committed predictions are recorded in the analysis log, which is available as stated in the data availability statement.

The exploratory samples of \cref{sec:diag8} were not blinded and have no prediction hash, the hashes of their received archives, and of the cp5a9 analysis script fixed before its events were generated, are recorded in the analysis log.

The inference code needs only \texttt{numpy}. Event generation uses \textsc{Pythia}~8.318 with the \texttt{TopRecon\allowbreak UserHooks} plugin, analysis and figures use Python with \texttt{numpy} and \texttt{matplotlib}.

The pre-registered analyses of \crefrange{sec:nlo}{sec:rivet}, with their shower and merge scripts and the extracted HEPData tables, were likewise hashed before their inputs existed. Every step, including the correction of \cref{sec:rivet-tau}, is recorded with time stamps in the analysis log.

\section{QCD background}
\label{app:theory}

This appendix complements \cref{sec:theory}.

\subsection{Colour flow and the timing of top decay}
\label{sec:theory-colour}

The top quark decays before it hadronises. With the width used by \textsc{Pythia}, $\Gamma_t=1.40\GeV$, the proper decay length is
\begin{equation}
  c\tau_t = \frac{\hbar c}{\Gamma_t} \approx \frac{0.197\GeV\,\mathrm{fm}}{1.40\GeV} \approx 0.14\,\mathrm{fm},
  \label{eq:ctau}
\end{equation}
well below the hadronisation length $\hbar c/\Lambda_{\mathrm{QCD}}\sim1\,\mathrm{fm}$. The $b$-quark therefore inherits the colour charge of the top and hadronises in the same space-time region as the beam remnants, the multiparton-interaction (MPI) systems and the other top's decay products. At leading colour, the $b$-quark is the endpoint of a colour string whose other end lies on a parton from the initial state, a beam remnant or radiation, depending on the production channel and the shower history. Gluons radiated in the decay sit on the same string.

In the Lund string model the hadron multiplicity grows with the string length, conveniently measured by
\begin{equation}
  \lambda = \sum_{\text{dipoles } ij} \ln\!\Big(1+\frac{m_{ij}^2}{m_0^2}\Big), \qquad m_0^2 = 1\GeV^2,
  \label{eq:lambda}
\end{equation}
summed over colour-connected parton pairs. Colour reconnection (CR) models change which partons are connected and therefore change $\lambda$, the multiplicity and, through the string tension, the distribution of energy between the $b$-jet cone and the rest of the event. The models used here are:
\begin{itemize}
  \item \emph{MPI-based CR} (\textsc{Pythia} default, mode 0). An MPI system of hardness $\pT$ is merged with a harder system with probability $P=p_{\mathrm{T0,rec}}^2/(p_{\mathrm{T0,rec}}^2+\pT^2)$, where $p_{\mathrm{T0,rec}}$ is the range parameter (default 1.8) times the MPI regularisation scale, partons of the softer system are inserted where they minimise $\lambda$. With late resonance decays (\texttt{PartonLevel:earlyResDec = off}) the top decay products are added after this step and are not themselves reconnected.
  \item \emph{QCD-based CR} (mode 1)~\cite{Christiansen:2015yqa}, which uses SU(3) colour algebra beyond leading colour, allows junction formation, and minimises $\lambda$ subject to causality constraints.
  \item \emph{Gluon move} (mode 2), which moves gluons between strings when this lowers $\lambda$.
  \item \emph{Top-specificCR}~\cite{Argyropoulos:2014zoa}, implemented as \texttt{TopReconUserHooks}, which acts on the gluons of the $t\to bW$ systems after the decay. Each such gluon is selected with probability $s$, the strength, and its colour and anticolour are swapped with those of a gluon from the rest of the event. The partner is chosen at random (mode 1), as the one with the smallest (mode 2) or largest (mode 3) invariant mass with the top gluon, as the one giving the smallest change in $\lambda$ (mode 4), or only if $\lambda$ decreases (mode 5).
\end{itemize}
Mode 3 reconnects each selected gluon to its most distant partner in invariant mass and is therefore the most disruptive. With $s$ continuous it defines the one-parameter family used for inference in \cref{sec:inference}.

\subsection{Heavy-quark fragmentation}
\label{sec:theory-frag}

The $b$-quark radiates less than a light quark at small angles. Emission off a quark of mass $m_Q$ and energy $E$ is suppressed inside the dead cone $\theta\lesssim\theta_0=m_Q/E$~\cite{Dokshitzer:1991fd}, for $m_b=4.8\GeV$ and the top-rest-frame energy $E_b=(\mt^2-\mW^2)/(2\mt)=67.5\GeV$, $\theta_0\approx0.07$. The $b$-quark keeps a large fraction of its energy, and the leading $B$ hadron takes a hard share of the jet.

That share is set by the fragmentation function $f(z)$, the probability for a hadron to take a fraction $z$ of the available light-cone momentum. In the Lund model~\cite{Andersson:1983ia} with the Bowler modification for heavy quarks~\cite{Bowler:1981sb},
\begin{equation}
  f(z) \propto \frac{1}{z^{1+r_b\, b\, m_b^2}}\,(1-z)^{a}\,\exp\!\Big(-\frac{b\,m_\perp^2}{z}\Big),
  \label{eq:bowler}
\end{equation}
with $a=0.68$, $b=0.98\GeV^{-2}$ and $r_b=0.855$ in the Monash tune. Larger $r_b$ enhances small $z$ and softens the fragmentation. The Peterson form~\cite{Peterson:1982ak},
\begin{equation}
  f(z) \propto \frac{1}{z\,\big(1-1/z-\varepsilon_b/(1-z)\big)^2}, \qquad \varepsilon_b = 0.005,
  \label{eq:peterson}
\end{equation}
is a common alternative. Both act at fixed $b$-quark energy: they redistribute that energy among the hadrons of the jet, but move little of it out of the cone. This is the physical difference from CR exploited below, within the variations considered here, the recoil scheme of the top decay, studied in \cref{sec:recoil}, also moves energy across the cone boundary. A softer fragmentation and CR both produce more, softer particles in the cone, so at the level of means they are nearly degenerate (\cref{sec:fingerprint}). Among the variations of this section, only CR removes charged energy from the cone and adds particles to the UE.

\subsection{The in-situ principle and identifiability}
\label{sec:theory-fisher}

The idea parallels the in-situ jet energy scale in lepton+jets top-mass measurements, where the hadronic $W$ mass constrains the light-jet response in the same events. Here the $b$-jet charged-particle flow constrains the CR strength $s$, and through it the CR-induced $b$-jet response. Linearising the model of \cref{eq:model} around a point $s_0$, the expected observables depend on $\theta=(\Delta\mt, s, \nu_j)$ through the Jacobian $J=(g,\,b,\,n_j)$. With the covariance held fixed at $C(s_0)$, the Fisher information is
\begin{equation}
  I(\theta) = J^{\top} C^{-1}(s_0)\, J + \mathrm{diag}(0,0,1,\dots,1),
  \label{eq:fisher}
\end{equation}
where the last term encodes the $\mathcal{N}(0,1)$ constraints on the nuisances. Since the covariance of \cref{eq:cov} depends on $s$, the complete Fisher matrix of the Gaussian likelihood contains in addition $\tfrac12\,\mathrm{Tr}(C^{-1}\partial_aC\,C^{-1}\partial_bC)$, \cref{eq:fisher} is the fixed-covariance (mean-response) approximation used for the estimates below, while the frozen likelihood itself uses the full $C(s)$. In this approximation $\Delta\mt$ and $s$ are separately identifiable if, and only if, $g$ and $b$ are not parallel in the metric $C^{-1}$ after the nuisance directions are projected out.

With the mass alone, $J$ has a single row: $\Delta\mt$ and $s$ enter only through the combination $\Delta\mt+b_m s$, with $b_m=-4.8\GeV$ per unit $s$. The mass is then determined only up to $b_m\,\Delta s=0.72\GeV$ over the family, which is why a mass-only method needs the envelope of \cref{sec:val24}.

With the frozen model at $s_0=0.075$, \cref{eq:fisher} gives
\begin{itemize}
  \item $\sigma(\Delta\mt)=0.102\GeV$ with $s$ and the nuisances profiled, against $0.097\GeV$ if $s$ were known,
  \item $\sigma(s)=0.0036$, so the trained range $[0,0.15]$ spans about 40 standard deviations,
  \item a correlation of $+0.31$ between $\Delta\mt$ and $s$, and a cosine of $-0.30$ between $g$ and $b$ in the whitened space.
\end{itemize}
Profiling the CR strength therefore costs only 5\% in mass resolution. The Fisher width agrees with the median 68\% half-width of $0.101\GeV$ found in the blinded validation (\cref{tab:val24}), a non-trivial check that the frozen covariance describes the actual sample-to-sample scatter.

The sensitivity to $s$ is spread over the flow features. Per $\Delta s=0.15$, the shifts in units of the single-sample standard deviation are $+38$ for $\langle\NUE\rangle$, $+21$ for $\langle\SUE\rangle$, $+18$ for $f(\Ninb\ge26)$, $+15$ for $\langle\Ninb\rangle$, $-16$ for $\langle\ptinb\rangle$, between $-8$ and $-15$ for the $\ptinb$ quantiles, and $-7.5$ for the fitted mass. A $1\GeV$ change of $\mt$ moves the fitted mass by 10.5 standard deviations but no flow feature by more than 3.

The method rests on one assumption: the true model lies on the one-dimensional curve in the ten-dimensional observable space traced out by the trained family. Components of the residual orthogonal to that curve are not absorbed by any parameter and inflate $\chimin$. This is what makes the goodness-of-fit statistic a test of the family assumption, and it explains why models with a different response pattern, such as CR off, are detected while models that mimic the top-specific pattern are not (\cref{sec:oof12}).

\section{Generator settings}
\label{app:generator}

Settings of the original leading-order Monash sample grid (\crefrange{sec:setup}{sec:oof12}), the CP5, NLO and 13\,TeV samples are described in \cref{sec:cms7,sec:diag8,sec:nlo,sec:rivet}: \texttt{Beams:eCM = 13600}, \texttt{Top:gg2ttbar = on}, \texttt{Top:qqbar2ttbar = on}, \texttt{24:onMode = off}, \texttt{24:onIfAny = 11 13}, Monash 2013 tune, 700\,000 events per sample. \Cref{tab:settings} lists the settings changed per model, as written to the \textsc{Pythia} log.

\begin{table*}[tbp]
  \centering
  \caption{\textsc{Pythia}~8.318 settings per model tag (tags as in \cref{tab:models}). $\varepsilon_b$: Peterson fragmentation parameter for $b$ quarks.}
  \label{tab:settings}
  \footnotesize
  \begin{tabular}{l>{\raggedright\arraybackslash}p{0.68\textwidth}}
    \toprule
    Tag & Settings \\
    \midrule
    \mtag{m0\_late} (nominal) & \texttt{ColourReconnection:mode = 0}, \texttt{PartonLevel:earlyResDec = off} \\
    \mtag{m0\_early} & \texttt{ColourReconnection:mode = 0}, \texttt{PartonLevel:earlyResDec = on} \\
    \mtag{m0\_rangeLo/Hi} & \texttt{ColourReconnection:range = 1.2} / \texttt{2.4} \\
    \mtag{m0\_pt215} & \texttt{MultipartonInteractions:pT0Ref = 2.15} \\
    \mtag{m1} & \texttt{ColourReconnection:mode = 1}, \texttt{BeamRemnants:remnantMode = 1}, \texttt{ColourReconnection:allowDoubleJunRem = off}, \texttt{MultipartonInteractions:pT0Ref = 2.15} \\
    \mtag{m2} & \texttt{ColourReconnection:mode = 2} \\
    \mtag{cr\_off} & \texttt{ColourReconnection:reconnect = off} \\
    \mtag{top1}--\mtag{top5} & nominal + \texttt{TopRecon:mode = 1}--\texttt{5}, \texttt{TopRecon:strength = 0.075} (mode 5: 1.0) \\
    \mtag{top3s<s>} & nominal + \texttt{TopRecon:mode = 3}, \texttt{TopRecon:strength =} $s$ \\
    \mtag{bfrag\_lo/hi} & \texttt{StringZ:rFactB = 0.70} / \texttt{1.00} \\
    \mtag{bfrag\_pet} & \texttt{StringZ:usePetersonB = on} (default $\varepsilon_b$) \\
    \mtag{frag\_aLund} & \texttt{StringZ:aLund = 0.78} (default 0.68) \\
    \bottomrule
  \end{tabular}
\end{table*}

\texttt{StringZ:useOldAExtra = on} appears among the changed settings in all 61 generator logs. It is set by the default Monash $e^+e^-$ tune file (\texttt{Monash2013-ee.cmnd}), because the tune was fitted with the pre-8.315 handling of the diquark and strange-quark $a$ parameters~\cite{Pythia8318doc}. It is therefore common to every sample and does not distinguish \mtag{bfrag\_pet}. According to the \textsc{Pythia} documentation, the corrected handling mainly changes baryon--antibaryon correlations and affects event shapes only mildly.

\paragraph{CMS tunes (cms7)} The CMS samples use \textsc{Pythia}~8.318 built against LHAPDF~6.5.5 and the CP5, CP5-CR1 and CP5-CR2 settings of the CMS software for 13.6\,TeV (the Run-3 tune configuration files~\cite{CMSSW:tunes}). The parameter values are copied unchanged, the tunes were derived with earlier \textsc{Pythia} versions, so two points need care.
\begin{itemize}
  \item \emph{CP5-CR1 in Pythia~8.311 and later.} Version 8.311 split the former \texttt{ColourReconnection:m0} into \texttt{m0} and a new pseudo-particle cut \texttt{mPseudo}. CP5-CR1 was fitted when \texttt{m0} played both roles, so we set \texttt{ColourReconnection:mPseudo = 1.721}, equal to \texttt{m0}. The same release changed parts of junction fragmentation and the QCD-based CR code, so CP5-CR1 in 8.318 is the closest available implementation, not an exact reproduction.
  \item \emph{CP5-CR2 parameters.} Five tuned values in the CMS software differ from the published table of Ref.~\cite{CMS:2022awf} (\cref{tab:cr2}). The 2017 production configuration agrees with the Run-3 file, so we follow the CMS software, which is what CMS generates with.
\end{itemize}

\begin{table*}[tbp]
  \centering
  \caption{CP5-CR2 parameters in the CMS software and in the published table.}
  \label{tab:cr2}
  \small
  \begin{tabular}{lcc}
    \toprule
    Parameter & CMS software (used) & Published table \\
    \midrule
    \texttt{MultipartonInteractions:pT0Ref} & 1.454 & 1.454 \\
    \texttt{MultipartonInteractions:ecmPow} & 0.0555 & 0.054 \\
    \texttt{MultipartonInteractions:coreFraction} & 0.4392 & 0.489 \\
    \texttt{MultipartonInteractions:coreRadius} & 0.6532 & 0.649 \\
    \texttt{ColourReconnection:m2Lambda} & 4.395 & 4.917 \\
    \texttt{ColourReconnection:fracGluon} & 0.9896 & 0.993 \\
    \bottomrule
  \end{tabular}
\end{table*}

\section{Origin of the generator-tune dependence}
\label{app:diag8}\label{sec:diag8}

This appendix is exploratory and is summarised in \cref{sec:cms7-origin}. Its two studies were generated after cms7 had been unblinded, and nothing in them changes the frozen method or the score of any blind test, in particular, the cms7 result of \cref{sec:cms7} stands. The first study (diag8) asks what moves the $\mlb$ fit between Monash and CP5. The second (cp5a9) builds the calibration on CP5 and measures the mass and flow responses within it, its analysis was fixed and hashed before its events were generated.

\subsection{Decomposition of the Monash-to-CP5 shift}
\label{sec:diag8a}

\paragraph{Design} Eight samples at $\mt=172.5\GeV$, each with $7\times10^5$ events and independent seeds: a fresh Monash sample, CP5, CP5-CR1, CP5-CR2, and four Monash samples that each carry one block of the 19 settings in which CP5 differs from Monash. Block A is the final-state shower coupling ($\alpha_s=0.1365$ with one-loop running $\to0.118$ with two-loop running), B is the PDF (NNPDF3.1 NNLO), C is the underlying-event block (the MPI parameters and coupling, the CR range and the total cross section, 12 settings), D is the initial-state shower and hard-process coupling (4 settings). Generated with the same seed, the union of the four blocks reproduces CP5 exactly, in its settings and event by event. Each $b$-jet also records its transverse momentum, the momentum fraction $z_B$ of its leading $B$ hadron along the jet axis, and the response $R_b^{\rm jet}=p_T^{\text{jet}}/p_T^{b}$ (not to be confused with the Bowler parameter $r_b$), where $b$ is the quark from the top decay matched within $\Delta R<0.4$, the 2.8--3.6\% of jets without a match are excluded from $R_b^{\rm jet}$.

\begin{table*}[tbp]
  \centering
  \caption{Exploratory diagnosis (ISR: initial-state radiation) at $\mt=172.5\GeV$ (one sample of $7\times10^5$ events each). Shifts are relative to the fresh Monash sample, the mass-only fit uses the frozen Monash templates. Statistical uncertainties: $\pm0.13\GeV$ on each mass shift, $\pm0.12\%$ on each response change, $\pm0.09$ on each $\NUE$ shift. $\delta\langle R_b^{\rm jet}\rangle$: change of the mean $b$-jet response $p_T^{\rm jet}/p_T^{b}$, $\NUE$: charged multiplicity outside jets, $\chimin$: goodness-of-fit statistic of the frozen method. All seven samples exceed its threshold.}
  \label{tab:diag8}
  \small
  \begin{tabular}{llcccr}
    \toprule
    Sample & Settings changed from Monash & Mass-only shift & $\delta\langle R_b^{\rm jet}\rangle$ & $\delta\langle\NUE\rangle$ & $\chimin$ \\
     & & [GeV] & [\%] & & \\
    \midrule
    A: FSR coupling & 2 & $+0.54$ & $+0.64$ & $-2.1$ & 131 \\
    B: PDF & 1 & $-0.74$ & $-0.98$ & $-15.1$ & 553 \\
    C: UE block & 12 & $+1.81$ & $+1.95$ & $+30.7$ & 34341 \\
    D: ISR coupling & 4 & $-0.28$ & $-0.35$ & $-2.3$ & 77 \\
    \midrule
    CP5 & 19 (all of A--D) & $+0.67$ & $+0.52$ & $+1.8$ & 631 \\
    CP5-CR1 & 28 & $+0.90$ & $+0.49$ & $+2.7$ & 3849 \\
    CP5-CR2 & 21 & $+0.85$ & $+0.45$ & $+2.8$ & 5767 \\
    \bottomrule
  \end{tabular}
\end{table*}

\paragraph{Mass shift and $b$-jet response} \Cref{tab:diag8} gives the results. Changing only the final-state coupling shifts the fit by $+0.54\pm0.13\GeV$, comparable to the $+0.67\pm0.13\GeV$ of full CP5. This does not make the decomposition unique. The PDF block alone gives $-0.74\GeV$ and the underlying-event block alone $+1.81\GeV$, and the four blocks sum to $+1.33\GeV$, not $+0.67\GeV$. Blocks B and C act mainly through the level of the underlying event: CMS fitted its MPI parameters together with its PDF, so each block on its own moves $\langle\NUE\rangle$ (about 45 in Monash) by $-15$ or $+31$ charged particles, far from any tuned value, while in full CP5 the two nearly cancel. The common thread is the $b$-jet response. Across the four single blocks the mass shift is $0.89\GeV$ per 1\% change in $\langle R_b^{\rm jet}\rangle$, close to the $0.74\GeV$ per 1\% expected from \cref{eq:jes,eq:dm-estimate}: a smaller final-state coupling lets less radiation leave the $R=0.4$ cone, while the PDF and underlying-event blocks change how much underlying-event energy falls into it (\cref{fig:diag8}, left). For the three complete CMS tunes the shift is larger than the response alone predicts, by about $+0.2$--$0.5\GeV$, with one sample each this residual is not significant.

\paragraph{The CMS tunes at matched mass} Combining the three samples here with the six cms7 samples, the mass-only fit relative to the frozen Monash reference is shifted by $+0.54\pm0.06\GeV$ for CP5, $+0.60\pm0.06\GeV$ for CP5-CR1 and $+0.57\pm0.06\GeV$ for CP5-CR2. All nine samples are consistent with one common shift of $+0.57\pm0.03\GeV$ ($\chi^2=8.0$ for 8 degrees of freedom), and the CR variants differ from CP5 by $+0.05\pm0.07\GeV$ (with the template statistics included and CP5 templates, $+0.05\pm0.05\GeV$, \cref{sec:cp5a9}). These are differences between complete tune configurations, not isolated effects of the CR algorithms: CP5-CR1 and CP5-CR2 carry their own MPI parameters, and CP5-CR1 its own fragmentation parameters. The uncertainties here are statistical. The frozen reference is common to all samples and cancels in the differences, and the cms7 samples at 172.0 and $173.0\GeV$ assume a unit response of the fit to the mass. The fresh Monash sample lies $0.23\pm0.12\GeV$ below the frozen reference, so shifts quoted against it (\cref{tab:diag8}) are larger by that amount. The three tunes give the same $b$-jet response to within 0.1\%. CP5-CR1 lowers $\langle z_B\rangle$ by 1.3\% through its softer string fragmentation, but not the jet response: the jet collects the fragmentation products whatever their momentum sharing.

\paragraph{Flow without a universal mass relation} At matched mass the flow separates CP5-CR1 and CP5-CR2 from CP5 clearly: the nine features give $\chi^2=2285$ and 3586 for one sample each, with the Monash covariance. Projected onto the trained top-specific CR direction, the flow changes look like $s\approx0.13$ and 0.15 (whitened cosines $+0.87$ and $+0.84$). The trained relation would then imply mass shifts of about $-0.6$ and $-0.7\GeV$, the observed shifts are consistent with zero. Full CP5 relative to Monash behaves in the same way: its flow change resembles $s\approx0.10$ (cosine $+0.68$), which would imply $-0.50\GeV$, while the fit moves by $+0.67\GeV$. This is why the in-situ correction of the frozen method made the CMS samples worse rather than better, moving the CP5 sample here from $+0.44$ (mass only) to $+1.44\GeV$ (\cref{fig:diag8}, right).

\begin{figure*}[tbp]
  \centering
  \includegraphics[width=\textwidth]{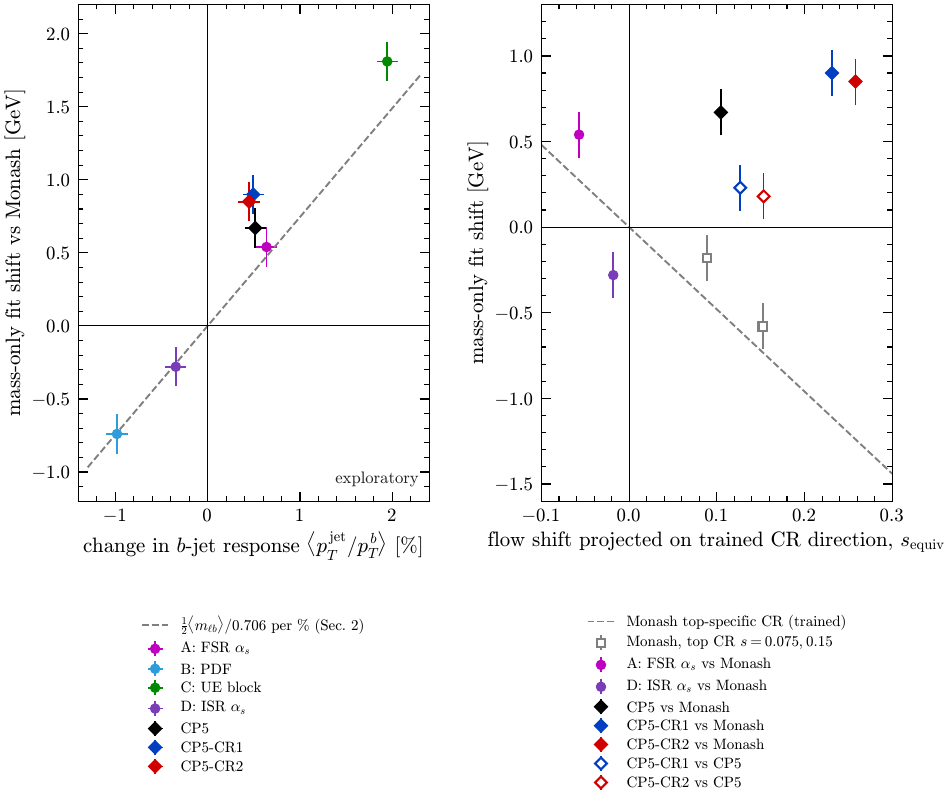}
  \caption{Exploratory diagnosis of the Monash-to-CP5 shift. Left: mass-only fit shift against the change in $b$-jet response for the four single blocks of CP5 (circles) and the three complete CMS tunes (diamonds), the dashed line is the expectation from \cref{eq:jes,eq:dm-estimate}. Right: mass-only shift against the flow change projected on the trained top-specific CR direction, expressed as an equivalent strength $s_{\rm equiv}$. The dashed line is the trained Monash relation, open squares are Monash top-specific CR samples. Filled diamonds compare the CMS tunes with Monash, open diamonds compare CP5-CR1 and CP5-CR2 with CP5. Shifts are relative to the fresh Monash sample, which lies $0.23\GeV$ below the frozen reference, so Monash points sit about $0.2\GeV$ above the line. Error bars are statistical.}
  \label{fig:diag8}
\end{figure*}

\subsection{Calibration and responses within CP5}
\label{sec:cp5a9}

\paragraph{Design} Nine further samples with CP5, each with $7\times10^5$ events and independent seeds: $\mlb$ templates at $\mt=171.5$, 172.0, 173.0 and $173.5\GeV$, two replicas at $172.5\GeV$, top-specific CR (mode 3) on top of CP5 at $s=0.075$ and 0.15, and CP5 with the Monash final-state coupling. A smoke test required the new generator to reproduce the previous ones event by event, CP5 with the CR hook attached but inactive to reproduce CP5 exactly, zero reconnections at $s=0$, and every sample to differ from CP5 only by its intended settings, no sample had a generator failure. The analysis uses the frozen v3 observables with CP5 templates, an event bootstrap for the statistical covariance and Poisson replicas of the templates, recomputed in every difference so that the shared template uncertainty is propagated. A second version of the analysis, also fixed before the data, adds the correlations from the shared CP5 baseline, both give the same conclusions, and the numbers below are from the second. CP5, CP5-CR1 and CP5-CR2 samples from diag8 and cms7 are reused.

\begin{table*}[tbp]
  \centering
  \caption{Results within CP5 (CP5 templates). Closure: fit minus $(\mt-172.5\GeV)$ on five independent CP5 samples. Shifts: relative to the mean of three CP5 replicas at $172.5\GeV$. Uncertainties combine statistics and template statistics. ``Reverse'': the opposite change, Monash with the CP5 FSR coupling (\cref{tab:diag8}).}
  \label{tab:cp5a9}
  \footnotesize
  \begin{tabular}{lcc}
    \toprule
    Quantity & Value [GeV] & Remark \\
    \midrule
    Closure, CP5 templates (5 samples) & $-0.02\pm0.06$ & $\chi^2=4.4$ for 4 d.o.f. \\
    Closure, Monash templates (same samples) & $+0.53\pm0.04$ & CMS-tune offset \\
    \midrule
    Top-specific CR, $s=0.075$ & $-0.23\pm0.10$ & \\
    Top-specific CR, $s=0.15$ & $-0.50\pm0.11$ & $4.7\sigma$ \\
    Slope $\mathrm{d}\hat m/\mathrm{d}s$ & $-3.3\pm0.7$ per unit $s$ & Monash training: $-4.8$ \\
    CP5 with Monash FSR coupling & $-0.49\pm0.10$ & reverse: $+0.54\pm0.13$ \\
    CP5-CR1 and CP5-CR2 minus CP5 (6 vs 5 samples) & $+0.05\pm0.05$ & 95\% range $[-0.05,+0.16]$ \\
    \bottomrule
  \end{tabular}
\end{table*}

\paragraph{Calibration} With CP5 templates the five independent CP5 samples, at 172.0 and $172.5\GeV$ and from three generation campaigns, close to $-0.02\pm0.06\GeV$, with a normal spread (\cref{tab:cp5a9}, \cref{fig:cp5a9}). The same samples fitted with the frozen Monash templates are shifted by $+0.53\GeV$. The average tune-dependent offset behind the cms7 failure is therefore a calibration effect, and it is removed by building the templates in the tune used.

\paragraph{Responses} Top-specific CR still lowers the fitted mass on CP5, by $-0.50\pm0.11\GeV$ at $s=0.15$, but the slope, $-3.3\pm0.7\GeV$ per unit strength, is about 30\% smaller than in the Monash training family. The mass correction for a given CR strength is therefore tune dependent. Restoring the Monash final-state coupling within CP5 moves the fit by $-0.49\pm0.10\GeV$, the mirror image of the $+0.54\pm0.13\GeV$ obtained by giving Monash the CP5 coupling (\cref{sec:diag8a}): the final-state coupling moves the fit by about half a GeV whichever baseline it is varied on. The complete CMS CR variants change the fit by $+0.05\pm0.05\GeV$ relative to CP5.

\paragraph{Directions} In the flow features, whitened with the CP5 covariance, top-specific CR and the final-state coupling point in clearly different directions (cosine $+0.43\pm0.03$, a proportionality test gives $\chi^2=1493$ for 8 degrees of freedom). The CP5-CR1 and CP5-CR2 responses are highly aligned with each other within this observable basis (cosine $+0.99$) and with top-specific CR (cosines $+0.90$ and $+0.87$), yet not proportional to it at the present statistics ($\chi^2=444$ and 661 for 8 degrees of freedom, 148 and 113 when the FSR direction is also allowed). A preliminary linear model in $(\Delta\mt,s,f_{\rm FSR})$, with response vectors taken from single samples, recovers the mass of independent CP5 samples ($-0.42\pm0.09\GeV$ for a true $-0.5\GeV$) with a precision of $0.081\GeV$ per $7\times10^5$ events, almost the $0.079\GeV$ obtained if $s$ and $f_{\rm FSR}$ were known. Applied to CP5-CR1 and CP5-CR2, which are outside the model, it shifts the mass by $+0.1$ to $+0.3\GeV$, but its goodness of fit rejects every one of them ($\chi^2=300$--370 for 7 degrees of freedom).

\begin{figure*}[tbp]
  \centering
  \includegraphics[width=\textwidth]{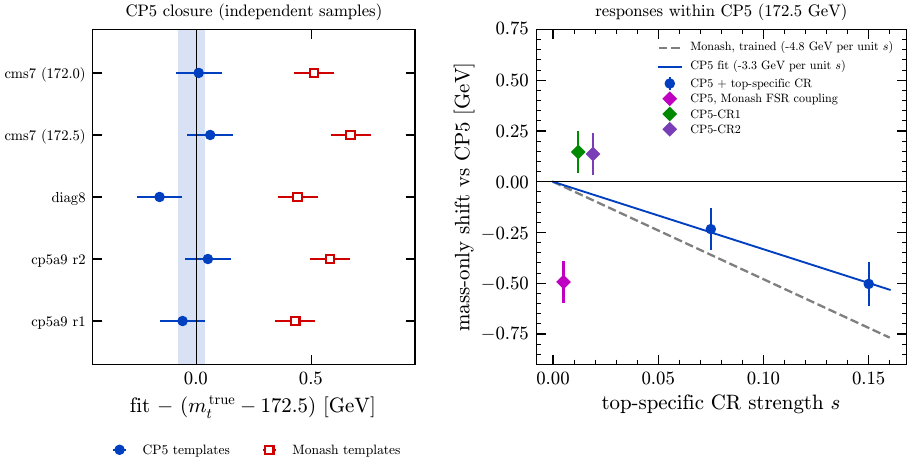}
  \caption{Calibration and responses within CP5 (exploratory). Left: fitted minus true mass for five independent CP5 samples with CP5 templates (filled, band: mean $\pm$ uncertainty) and with the frozen Monash templates (open). Right: mass-only shift relative to CP5 at $172.5\GeV$ as a function of the top-specific CR strength (circles, solid line: fitted slope), compared with the trained Monash relation (dashed), CP5 with the Monash FSR coupling and the complete CMS CR variants are shown at the left edge. Error bars include template statistics.}
  \label{fig:cp5a9}
\end{figure*}

\subsection{Implications for inference}
\label{sec:implications}

The charged-particle flow responds strongly to changes of colour reconnection and underlying-event modelling without a universal relation to the corresponding shift of the top mass. Top-specific CR moves both the flow and the mass. Complete CMS tune variations move the flow in a similar direction and leave the mass nearly unchanged. A flow-based mass correction is therefore valid only within a model family whose flow-to-mass relation has been established, and its calibration depends on the tune.

The CP5 results show how such a model could be built: templates and response vectors on CP5, separate nuisance directions at least for the final-state shower and for top-specific CR, and complete alternative tunes treated as discrete variants that the goodness-of-fit test must flag, with a mass-only fallback whose bias for those variants is small. None of this has yet been tested blind. The next-to-leading-order check of the responses is the subject of \cref{sec:nlo}, a frozen CP5 model with a new prospective test on independent samples is left for future work (\cref{sec:outlook}).

\section{Covariance checks of the HEPData record}
\label{app:hepdata}

\begin{table*}[tbp]
  \centering
  \caption{Checks of the normalised covariance matrices of HEPData record ins2152933 v1. ``Theory'': including the theory uncertainties of the measurement.}
  \label{tab:hepdata}
  \footnotesize
  \begin{tabular}{p{0.30\textwidth}lcp{0.34\textwidth}}
    \toprule
    Observable & Tables & Status & Problem \\
    \midrule
    $n_{\rm ch}$, theory & 26 & inconsistent & asymmetric entries, negative eigenvalue, normalisation \\
    $n_{\rm ch}$ and $\sum\pT$ (global), theory & 53--56 & inconsistent & entries shifted, normalisation violated \\
    $\sum\pT$ in $n_{\rm ch}$ slices, theory & 28--52 & inconsistent & asymmetric (1\%), normalisation (3\%) \\
    slices, block $60$--$80\times40$--$60$ & 45, 76 & mislabelled & replaced by the transpose of 41, 72 \\
    $n_{\rm ch}$ without theory & 57 & valid & primary \\
    $\sum\pT$ (theory / without) & 27 / 58 & valid & cross-check \\
    global and slices without theory & 84--87, 59--83 & valid & cross-check \\
    \bottomrule
  \end{tabular}
\end{table*}

The checks of \cref{sec:rivet-hepdata} use HEPData record ins2152933, version~1, read without transformation from the published YAML tables. Each table is placed by its own labels (variable names, and the $n_{\rm ch}$ slice ranges given in its description), not by its table number. \Cref{tab:hepdata-entries} lists the inconsistent entries of the $n_{\rm ch}$ covariance including theory uncertainties (table 26), values are in the units of the record.

\begin{table}[tbp]
  \centering
  \caption{Mirror entries of HEPData table 26 ($n_{\rm ch}$, normalised, including theory uncertainties) that disagree. A symmetric covariance requires $C_{ij}=C_{ji}$.}
  \label{tab:hepdata-entries}
  \footnotesize
  \resizebox{\columnwidth}{!}{\begin{tabular}{lccl}
    \toprule
    Bins $(i,j)$ in $n_{\rm ch}$ & $C_{ij}$ & $C_{ji}$ & Apparent problem \\
    \midrule
    $[20,30),[30,40)$ & $2.6283846\times10^{-7}$ & $2.6283846\times10^{-8}$ & exponent differs by one \\
    $[20,30),[40,50)$ & $-7.4343859\times10^{-10}$ & $-7.4343859\times10^{-7}$ & exponent differs by three \\
    $[30,40),[40,50)$ & $-7.4343859\times10^{-10}$ & $1.3534734\times10^{-7}$ & copied from a neighbouring entry \\
    \bottomrule
  \end{tabular}}
\end{table}

After symmetrisation, table 26 has a smallest eigenvalue of $-2.0\times10^{-7}$ ($-1.9\times10^{-7}$ with the last bin removed). As published, two of its rows violate the normalisation constraint $\sum_jC_{ij}\Delta_j=0$ by 36\% and 71\% of the largest $C_{ii}\Delta_i$, after symmetrisation three rows violate it, by 18--47\%. In the $n_{\rm ch}$ block of the global covariance (table 53), 19 of the 28 mirror pairs disagree, in a pattern consistent with entries shifted by one position, its $[40,50)$ diagonal entry ($1.3527359\times10^{-7}$) equals an off-diagonal entry, while table 26 gives $2.0750996\times10^{-7}$, the smallest eigenvalue of the symmetrised $19\times19$ global matrix is $-4.4\times10^{-7}$ ($-3.7\times10^{-7}$ with the last bin of each distribution removed). The double-differential covariance including theory uncertainties (tables 28--52) remains asymmetric at the 1\% level and violates the normalisation by 3\% as published (2\% after symmetrisation), after the repair of table 45. Before that repair it also has a negative eigenvalue ($-5\times10^{-12}$), after it, it is positive definite with its one redundant bin removed. Table 45 duplicates table 40 and table 76 carries the bins of a different block, both are replaced by the transposes of their correctly labelled mirror blocks (tables 41 and 72). All matrices without theory uncertainties, and the $\sum\pT$ matrix with them, are symmetric, positive definite after removing the last bin, and satisfy the normalisation to better than $10^{-5}$. The extraction and check scripts are archived with the analysis code. A second implementation of the checks, written independently of the analysis code, reproduces all of the findings above from the published YAML files, including the double-differential blocks (tables 28--52 and 59--83).

\FloatBarrier
\bibliographystyle{elsarticle-num}
\bibliography{references}

\end{document}